%% file: arXiv.tex
\documentclass{article}

\usepackage{microtype}
\usepackage{graphicx}
\usepackage{subfigure}
\usepackage{booktabs} %

\usepackage{hyperref}

\usepackage[accepted]{arXiv}

\usepackage{amsmath}
\usepackage{amssymb}
\usepackage{mathtools}
\usepackage{amsthm}

\usepackage[capitalize,noabbrev]{cleveref}

\theoremstyle{plain}
\newtheorem{theorem}{Theorem}[section]

\newtheorem{lemma}[theorem]{Lemma}
\newtheorem{corollary}[theorem]{Corollary}
\theoremstyle{definition}
\newtheorem{definition}[theorem]{Definition}

\theoremstyle{remark}
\newtheorem{remark}[theorem]{Remark}

\usepackage[textsize=tiny]{todonotes}

\usepackage{lipsum}
\usepackage{dsfont}
\usepackage{graphicx}
\usepackage{float}
\usepackage{mathtools}
\usepackage{mathrsfs}
\usepackage{enumitem}
\usepackage{xspace}
\usepackage{multirow}
\usepackage[export]{adjustbox}
\usepackage[framemethod=TikZ]{mdframed}
\usepackage[strings]{underscore}
\usepackage{esvect}

\input{macros}

\icmltitlerunning{Tight Data Access Bounds for Private Top-$k$ Selection\hfill\thepage}

\begin{document}

\twocolumn[
\icmltitle{Tight Data Access Bounds for Private Top-$k$ Selection}

\begin{icmlauthorlist}
\icmlauthor{\texorpdfstring{$\begin{array}{c} \textrm{Hao Wu} \\ \small \textrm{whw4@student.unimelb.edu.au} \\ \textrm{The University of Melbourne} \\
\textrm{Australia} \\ \end{array}$}{Lg}}{}
\icmlauthor{\texorpdfstring{$\begin{array}{c} \textrm{Olga Ohrimenko} \\ \small \textrm{oohrimenko@unimelb.edu.au} \\ \textrm{The University of Melbourne} \\
\textrm{Australia} \\
\end{array}$}{Lg}}{}
\icmlauthor{\texorpdfstring{$\begin{array}{c} \textrm{Anthony Wirth} \\ \small \textrm{awirth@unimelb.edu.au} \\ \textrm{The University of Melbourne} \\
\textrm{Australia} \\ \end{array}$}{Lg}}{}
\end{icmlauthorlist}

]

\printAffiliationsAndNotice{}  %

\input{abstract}

\input{introduction}

\input{problem-definition}
\input{preliminaries}

\input{top-k-algorithm}

\input{lower-bounds.tex}

\input{related-work}

\input{conclusion}

\input{acknowledgement}

\bibliography{reference}
\bibliographystyle{icml2023}

\newpage
\appendix
\onecolumn
\input{appendix}

\end{document}

%% file: macros.tex
\newtheorem{example}[theorem]{Example}

\newtheorem{fact}[theorem]{Fact}

\renewenvironment{quote}
  {\list{}{\rightmargin=0.3cm \leftmargin=0.3cm}%
   \item\relax}
  {\endlist}

\NewDocumentCommand\p{ m g }{
  \ensuremath{
    \IfNoValueTF{#2}
    {\Pr [ #1 ]}
    {\Pr_{#1}[#2]}
  }
}
\RenewDocumentCommand\P{ m g }{
  \ensuremath{
    \IfNoValueTF{#2}
    {\Pr \left[#1\right]}
    {\Pr_{#1}\left[#2\right]}
  }
}
\DeclareMathOperator*{\Expectation}{\mathbb{E}}
\NewDocumentCommand\E{ m g }{
  \ensuremath{
    \IfNoValueTF{#2}
    {\Expectation \left[#1\right]}
    {\Expectation_{#1}\left[#2\right]}
  }
}
\DeclareMathOperator*{\Variance}{\mathbb{V}\mathrm{ar}}
\NewDocumentCommand\Var{ m g }{
  \ensuremath{
    \IfNoValueTF{#2}
    {\Variance \left[#1\right]}
    {\Variance_{#1}\left[#2\right]}
  }
}

\NewDocumentCommand\DataList{ m g }{
  \ensuremath{
    \IfNoValueTF{#2}
    {L_{#1}}
    {L_{#1}[{#2}]}
  }
}

\NewDocumentCommand\GammaDist{ m g }{
  \ensuremath{
    \IfNoValueTF{#2}
    {\operatorname{\mathcal{G}amma}\left( {#1} \right)}
    {\operatorname{\mathcal{G}amma}\left( {#1}, {#2} \right)}
  }
}

\NewDocumentCommand\LapNoise{ m g }{
  \ensuremath{
    \IfNoValueTF{#2}
    {\operatorname{\mathbb{L}ap}\left( {#1} \right)}
    {\operatorname{\mathbb{L}ap}\left( {#1}, {#2} \right)}
  }
}

\NewDocumentCommand\GumbelNoise{ m g }{
  \ensuremath{
    \IfNoValueTF{#2}
    {\operatorname{\mathbb{G}umbel}\left( {#1} \right)}
    {\operatorname{\mathbb{G}umbel}\left( {#1}, {#2} \right)}
  }
}

\newcommand{\invertedIndex}[1]{\sigma_{#1}}
\newcommand{\uniOrderStat}[1]{U_{({#1})}}

\newcommand{\Data}[1]{{\vec{x}_{#1}}}

\newcommand{\DataSet}{\mathcal{X}}
\newcommand{\hist}{\vec{h}}

\newcommand{\DataDomain}{\mathcal{C}}
\newcommand{\AggregationFunt}{f}
\newcommand{\algoOracle}{\mathcal{A}_{oracle}}
\newcommand{\algoOrderStat}{\mathcal{A}_{ord\text{-}stat}}
\newcommand{\AccessCost}[2]{cost\left( {#1}, {#2} \right)}

\newcommand{\HighFreqSet}{S_{H}}
\newcommand{\LowFreqSet}{S_{L}}

\newcommand{\BetaDist}[2]{\operatorname{\mathcal{B}eta}\left( {#1}, {#2} \right)}
\newcommand{\BetaFunt}[2]{\mathrm{B}( {#1}, {#2})}
\newcommand{\GammaFunt}[1]{\Gamma( {#1} )}

\NewDocumentCommand\idx{g}{
  \ensuremath{
    \IfNoValueTF{#1}{ \mathsf{idx} }{ \mathsf{idx}_{#1} }
  }
}

\NewDocumentCommand\dyadicInterval{ g g }{
  \ensuremath{
    \IfNoValueTF{#2}
    { \IfNoValueTF{#1}{ \mathcal{I} }{ \mathcal{I}_{#1, * } } }
    { \mathcal{I}_{#1, #2} }
  }
}
\NewDocumentCommand\dyadicIntervalSet{g}{
  \ensuremath{
    \IfNoValueTF{#1}{ \textit{ISet} }{ \textit{ISet}[{#1}] }
  }
}

\NewDocumentCommand\partialsum{gg}{
  \ensuremath{
    \IfNoValueTF{#2}
    { S ({#1} ) }
    { S_{#1} ({#2}) }
  }
}

\newcommand{\indicator}[1]{\mathds{1}_{\left[#1\right]}}

\newcommand{\PAREN}[1]{{\left( {#1} \right)}}
\newcommand{\paren}[1]{{( {#1} )}}

\newcommand{\card}[1]{\left| {#1} \right|}
\newcommand{\set}[1]{\left\{ {#1} \right\}}
\newcommand{\eps}{\varepsilon}

\DeclareMathOperator*{\argmax}{arg\,max}

\def\algoTA{ \mathcal{A}_{ \textit{TA} } }
\def\algoPrivTA{ \mathcal{A}_{ \textit{PrivTA} } }

\newcommand{\cA}{\mathcal{A}}

\newcommand{\cD}{\mathcal{D}}
\newcommand{\cE}{\mathcal{E}}
\newcommand{\cF}{\mathcal{F}}
\newcommand{\cG}{\mathcal{G}}

\newcommand{\cJ}{\mathcal{J}}
\newcommand{\cH}{\mathcal{H}}

\newcommand{\cM}{\mathcal{M}}
\newcommand{\cO}{\mathcal{O}}

\newcommand{\cS}{\mathcal{S}}

\newcommand{\cU}{\mathcal{U}}

\newcommand{\cX}{\mathcal{X}}

\newcommand{\N}{\mathbb{N}}
\newcommand{\R}{\mathbb{R}}

\NewDocumentCommand\vecp{g}{
  \ensuremath{
    \IfNoValueTF{#1}{ \vec{p} }{ \vec{p}^{\,(#1)} }
  }
}

\newif\ifcomment
\commentfalse

\definecolor{DarkGreen}{rgb}{0.1,0.5,0.1}
\ifcomment
\newcommand{\tony}[1]{\textcolor{brown}{[TONY: #1]}}
\newcommand{\hao}[1]{\textcolor{blue}{[HAO: #1]}}

\newcommand{\olya}[1]{\textcolor{DarkGreen}{[Olya: #1]}}
\newcommand{\olyanew}[1]{\textcolor{purple}{[OlyaNew: #1]}}
\else

\makeatletter
\newcommand{\tony}[1]{%
  \@bsphack
  \@esphack
}
\newcommand{\hao}[1]{%
  \@bsphack
  \@esphack
}
\newcommand{\olya}[1]{%
  \@bsphack
  \@esphack
}
\newcommand{\olyanew}[1]{%
  \@bsphack
  \@esphack
}
\makeatother
\fi

%% file: abstract.tex
\begin{abstract}
    
    We study the top-$k$ selection problem under the differential privacy model:~$m$ items are rated according to votes of a set of clients.
    We consider a setting in which algorithms can retrieve data via a sequence of accesses, each either a random access or a sorted access; the goal is to minimize the total  number of data accesses.
    Our algorithm requires only $O(\sqrt{mk})$ expected accesses: to our knowledge, this is the first sublinear data-access upper bound for this problem.
        Our analysis also shows that the well-known exponential mechanism requires only $O(\sqrt{m})$ expected accesses.
    Accompanying this, we develop the first lower bounds for the problem, in three settings: only random accesses; only sorted accesses; a sequence of accesses of either kind.
    We show that, to avoid~$\Omega(m)$ access cost, supporting \emph{both} kinds of access is necessary, and that in this case
    our algorithm's access cost is optimal.
    
\end{abstract}

%% file: introduction.tex
\section{Introduction}
\label{sec: introduction}

We consider the differentially private top-$k$ selection problem;  there are~$m$ items to be rated according to~$n$ clients' votes.
Each client can either vote or not vote for each item,
and can vote for an unlimited number of items.
Since this data can be sensitive (e.g., visited websites, purchased items, or watched movies), the goal is to identify a set of~$k$ items with approximately the highest number of votes, while concealing the  votes of individual clients.

Private top-$k$ selection is a fundamental primitive and underlies a wide range of differentially private machine learning and data analytics tasks such as 
discovering frequent patterns from data~\citep{BLST10}, 
training wide neural networks~\citep{ZMH21}, 
tracking data streams~\citep{Cardoso022}, 
false discovery rate control in hypothesis testing~\citep{QiaoSZ21}, 
etc.

In recent years, a significant progress has been made towards understanding how accurate the algorithms for this problem can be.
For example,~\citep{BafnaU17, SteinkeU17} provide lower bounds for the problem in terms of sample complexity, which can be achieved by a number of existing algorithms~\citep{DurfeeR19, QiaoSZ21}.

Another line of research is devoted to improving the efficiency of the algorithms.
Early works such as the peeling solution~\citep{BLST10} need to iterate $k$~times over all items. 
The improved mechanisms~\citep{DurfeeR19, QiaoSZ21} iterate over each item only once.
Since~$k$ can be much smaller than~$m$, the research community remains interested in the following question:

\vspace{-2mm}
\begin{quote}
    Is there a private top-$k$ selection algorithm that accesses only a sublinear number of items?
\end{quote}
\vspace{-2mm}
Although it seems to be an unachievable target, it is possible to address this question by considering how items are accessed.
For example,~\citeauthor{DurfeeR19}~\citeyearpar{DurfeeR19} consider the setting where the data has been pre-processed and resides in an existing data analytics system, that can return the items in sorted order 
(which we refer to as \emph{sorted access} in our paper).
Their top-$k$ algorithm can make a sublinear number of accesses at a cost of potentially returning fewer than~$k$ items, while to guarantee that~$k$ items are returned, the number of retrieved items can be~$m$.
Since retrieving information from an existing system incurs corresponding query processing and communication cost, it is crucial to minimize the number of data accesses.

In this paper, we systematically investigate the minimum number of items an algorithm needs to evaluate (a.k.a.~\emph{access cost}) , in order to answer the private top-$k$ selection problem.
In addition to \emph{sorted access}, we also consider another common way of accessing items' data, i.e., the \emph{random access}, in which an algorithm can actively request the data of an arbitrary item.\footnote{Here \emph{random} carries the sense of Random Access Memory (RAM), rather than the outcome of a random process.}
Both types of accesses have been considered by previous literature for the non-private version of top-$k$ selection problem (see~\citeauthor{IlyasBS08}~\citeyearpar{IlyasBS08} for a comprehensive survey).

\begin{example}
    Consider the example of a movie ranking database.
    It can present the movies in sorted order, according to their ratings by the clients, or it can return directly the rating of a specific movie.
\end{example}

\vspace{-2mm}
\paragraph{Our Contributions.} 
Our results are threefold.
On the upper bound side,
\begin{itemize}[leftmargin=4.5mm, topsep=2pt, itemsep=2pt, partopsep=2pt, parsep=2pt]
    \item If the system supports both sorted access and random access, 
    we design an algorithm with expected access cost~$O \paren{\sqrt{mk}}$.
\end{itemize}
To our knowledge, this is the first asymptotically sublinear bound of the access cost for the private top-$k$ selection problem.
Our algorithm builds on existing works~\citep{DurfeeR19, QiaoSZ21} and inherits their error bounds, which are known to be asymptotically optimal~\citep{BafnaU17, SteinkeU17}.
    Additionally, since the exponential mechanism~\citep{McSherryT07}, a fundamental technique in differential privacy, can be formulated as a private top-$1$ selection algorithm~\citep{DurfeeR19}, our result implies the following corollary:
    \begin{itemize}[leftmargin=4.5mm, topsep=2pt, itemsep=2pt, partopsep=2pt, parsep=2pt]
        \item[-] If the system supports both sorted access and random access, 
        the exponential mechanism requires only $O(\sqrt{m})$ expected accesses.
    \end{itemize}

On the lower bound side,
\begin{itemize}[leftmargin=4.5mm, topsep=2pt, itemsep=2pt, partopsep=2pt, parsep=2pt]
    \item If the system supports either only sorted accesses or only random accesses, but not both, we show a lower bound of~$\Omega\PAREN{m}$.
    \item If the system supports both sorted accesses and random accesses, we show a lower bound of~$\Omega\paren{\sqrt{mk}}$. 
\end{itemize}
These statements are informal versions of Theorems~\ref{theorem: lower bound for random access},~\ref{theorem: lower bound for sorted access}, and~\ref{theorem: lower bound for sorted and random access}, which impose modest assumptions on the privacy guarantee, and relatively weak assumptions on the accuracy guarantee of the algorithms. %
They show that supporting sorted and random access to the items' data simultaneously is necessary to break the linear barrier, and the access cost of our algorithm is essentially optimal. 

\vspace{-2mm}
\paragraph{Organization.} 
Our paper is organized as follows. Section~\ref{sec: problem definition} introduces the problem formally. 
Section~\ref{sec: preliminaries} discusses the preliminaries for our algorithm. 
Section~\ref{sec: solution} introduces our algorithm and shows the upper bounds. 
Section~\ref{sec: lower bounds} presents the lower bounds for the problem.
Section~\ref{sec: related-work} discusses the related works.
Section~\ref{sec: summary} summarizes the paper.

%% file: problem-definition.tex
\section{Model Description}
\label{sec: problem definition}

Let~$\DataDomain \doteq \set{1, \ldots, m}$ be a set of~$m$ items, and~$\cU \doteq \set{1, \ldots, n}$ be a set of~$n$ clients.
Each client~$v \in \cU$ can cast at most one vote for each item, and can vote for an unlimited number of items.
Hence, client~$v$'s votes, denoted by~$\Data{v}$, can be viewed as a vector in~$\cD \doteq \set{0, 1}^m$, such that for each~$i \in \DataDomain$, ~$\Data{v}[i] = 1$ if~$v$ votes for item~$i$, where~$\Data{v}[i]$ is the~$i^{(th)}$ entry of~$\Data{v}$.
We regard the collection of voting vectors from all~$n$ clients as a dataset~$\DataSet = \set{ \Data{1}, \ldots, \Data{n} } \in \cD^n$.

For each item~$i \in \DataDomain$, let its score~$\hist[i] \doteq \sum_{v \in \cU} \Data{v}[i]$ be the number of clients that vote for~$i$.
The dataset~$\DataSet$ can be described by its histogram~$\hist \doteq \paren{ \hist[1], \ldots, \hist[m] } \in \N^m$. 
We also define~$\pi: \DataDomain \rightarrow \DataDomain$ to be a permutation that puts the entries of~$\hist$ in nonincreasing order\footnote{We break ties arbitrarily.}, s.t.,~$\hist[\pi(1)] \ge \cdots \ge \hist[\pi(m)]$. 

Our goal is to design a differentially private algorithm that returns a set~$S$ of~$k$ items with (approximately) largest scores, while minimizing 
its data  access cost.

In what follows, we discuss the privacy guarantee, the utility guarantee, the data access model of an algorithm formally.

\vspace{-2mm}
\paragraph{\bf Privacy Guarantee.}
We call two datasets~$\DataSet, \DataSet'$ neighboring, denoted by~$\DataSet \sim \DataSet'$, if they differ  in addition or deletion of one client vector, e.g.,~$\DataSet' = \DataSet \cup \set{\Data{n + 1}}$ or~$\DataSet' = \DataSet \setminus \set{\Data{v}}$ for some~$v \in [n]$.

Let~$\hist$ and~$\hist'$ be the histograms corresponding to~$\DataSet$ and~$\DataSet'$, respectively. 
It is easy to see that if~$\DataSet \sim \DataSet'$, then the score of each item can differ by at most~$1$, i.e.,~$\Vert \hist - \hist' \Vert_\infty \le 1$.
Hence, for every~$\hist, \hist' \in \N^m$, we also call them neighboring histograms, written as~$\hist \sim \hist'$, 
    if and only if~$\Vert \hist - \hist' \Vert_\infty \le 1$.

    Let~$\tbinom{[m]}{k}$ be the collection of all subsets of~$[m]$ of size~$k$, and~$\cA: \N^m \rightarrow \tbinom{[m]}{k}$ be a top-$k$ selection algorithm.
To protect the voting information of individual clients, we would like its output distributions to be similar for neighboring inputs, as defined thus.

\begin{definition}[$\paren{\eps, \delta}$-Private Algorithm~\citep{DR14}] \label{def: Differential Privacy}
    Given~$\eps, \delta > 0$, a randomized algorithm 
    $\cA: \N^m \rightarrow \binom{[m]}{k}$
    is called~$\paren{\eps, \delta}$-differentially private (DP),
    if for every~$\hist, \hist' \in \N^m$ such that~$\hist \sim \hist'$, 
    and 
    all $Z \subseteq \binom{[m]}{k}$,
    \begin{equation} \label{ineq: def private algo}
        \P{ \cA (\hist) \in Z } \le e^\eps \cdot \Pr [ \cA (\hist') \in Z ] + \delta\,.
    \end{equation}
\end{definition}

We call~$\eps$ and~$\delta$ the \emph{privacy parameters}.
Typically, it is required that~$\delta$ is cryptographically negligible, i.e.,  $\delta \le 1 / m^{\omega(1)}$~\citep{Vadhan17, DR14}. 
An algorithm~$\cA$ is also called~$\eps$-DP for short, if it is~$\paren{\eps, 0}$-DP.

\vspace{-2mm}
\paragraph{\bf Utility Guarantee.}
In line with previous research~\citep{BafnaU17, DurfeeR19}, we measure the error of an output~$S$ by the maximum amount by which~$\hist[ \pi(k) ]$ exceeds the score of any item in~$S$, defined formally as follows. 

\begin{definition}[$\paren{\alpha, k}$-Accuracy]
    Given a vector~$\vec{h}$, parameters~$k \in \N^+$, and~$\alpha \in \R^+$, 
    an output~$S \in \binom{[m]}{k}$
    is called~$\paren{\alpha, k}$-accurate, if for each~$i \in S$,~$\hist[i] \ge \hist[\pi(k)] - \alpha$. 
\end{definition}

\vspace{-3mm}
\paragraph{\bf Data Access.} 
We assume that the histogram~$\hist$ has been preprocessing by an existing data management system, and an algorithm~$\cA$ can access~$\hist$ only through the system.
We consider two access models that abstract common functionalities supported by a system: \emph{sorted access} and \emph{random access}.
Such access models have been widely accepted by the community for non-private top-$k$ selection problems (see~\citeauthor{IlyasBS08}~\citeyearpar{IlyasBS08} for a survey).

\emph{Sorted Access}.
Let~$\DataDomain_s$ be the set of items already returned by sorted access (initially,~$\DataDomain_s = \varnothing$).
When a new sorted-access request is submitted, the system returns an item-score pair~$(i, \hist[i])$, where~$i \in [m] \setminus \DataDomain_s$ has the largest score, i.e., $i = \argmax_{j \in [m] \setminus \DataDomain_s} \hist[ j ]$.
An alternative view is that the system returns~$\big( \pi(1), \hist[\pi(1)] \big), \big( \pi(2), \hist[\pi(2)] \big), \ldots$ in order, one tuple at a time.

\emph{Random Access}.
A request of random access consists of a reference~$i \in \DataDomain$ to an item.
In response, the system returns the corresponding item-score pair~$(i, \hist[i])$.
We emphasize that a random access does not imply that~$i$ must be a randomly chosen item.

\emph{Access Cost.} 
Given an algorithm~$\cA$ and a histogram~$\hist$, the \emph{access cost} of an algorithm on~$\hist$,~$\AccessCost{\cA}{\hist}$, is the total number of accesses -- either sorted or random -- to $\hist$. %
Note that this is an upper bound of distinct number of entries~$\cA$ learns from~$\hist$, as a random access may retrieve a previously encountered item-score pair. 

%% file: preliminaries.tex
\vspace{-2mm}
\section{Preliminaries}
\label{sec: preliminaries}
\vspace{-1mm}

In this section, we review two building blocks for constructing our algorithm: 
a state-of-the-art algorithm for non-private top-$k$ selection, specifically designed for aggregating data from multiple sources, and a framework of the existing one-shot 
algorithms for private top-$k$ selection.

\vspace{-2mm}
\subsection{\bf Threshold Algorithm}
\vspace{-1mm}

The threshold algorithm~\citep{FaginLN03} is a top-$k$ selection algorithm, when the information of an item needs to be aggregated from multiple resources.
In this scenario, there are~$m$ items, each associated with~$t$ attributes.
Without loss of generality, assume that each attribute is a real number.
Therefore, each item~$i$ can be represented by a vector~$\vec{y}_i \in \R^t$.
The score of item~$i$ is computed by a function~$\AggregationFunt: \R^t \rightarrow \R$, which is assumed to be \emph{monotone}, s.t., for each~$\vec{y}, \vec{y}' \in \R^t$, if~$\vec{y}[j] \le \vec{y}'[j], \forall j \in [t]$, then~$\AggregationFunt(\vec{y}) \le \AggregationFunt(\vec{y}')$.

\newcommand{\val}{\mathsf{val}}
The vectors~$\vec{y}_1, \ldots, \vec{y}_m$ do not reside in a single data management system, but distributed in~$t$ systems~$\DataList{1}, \ldots, \DataList{t}$, s.t., for each item~$i$, its~$j^{th}$  attribute~$\vec{y}_i[j]$ resides on~$\DataList{j}$.
Each~$\DataList{j}$ allows for both \emph{sorted access} and \emph{random access}.
We can view it as an array of~$m$ tuples~$\DataList{j}{1}, \ldots, \DataList{j}{m}$, each of the form~$(i, \val) \in [m] \times \R$, where~$\val$ equals~$\vec{y}_{i}[j]$.
The tuples in~$\DataList{j}$ are sorted in descending order by their~$\val$'s. 
Further,~$\DataList{j}$ is augmented with an inverted index~$\invertedIndex{j}: [m] \rightarrow [m]$ to support random access, such that, for each an item~$i \in [m]$,~$\DataList{j}[\invertedIndex{j}(i)]$ contains the tuple~$\paren{i, \vec{y}_{i}[j]}$.

The aim is to identify the top-$k$ items with highest scores according to~$f$, while minimizing the access cost, i.e., the total number of data accesses performed by the algorithm to $\DataList{1}, \ldots, \DataList{t}$.
The algorithm is described in Algorithm~\ref{algo: threshold algorithm}.

    It works in round-robin fashion.
    In each round, it retrieves one tuple from each sorted array~$\DataList{j}$.
    For each tuple~$(i, \val)$ encountered during sorted access, it retrieves all entries~$\vec{y}_{i}[j]$ of~$\vec{y}_{i}$ by random accesses. 
    This step can be optimized at a cost of memoization:
    we augment~$\algoTA$ with a data structure to store previously encountered~$i$'s.
    After retrieving all~$\vec{y}_{i}[j]$, the algorithm computes~$\AggregationFunt(\vec{y}_{i})$, and maintains a set~$S$, consisting of~$k$ item-score pairs with largest scores seen so far.
    The algorithm stops when there are~$k$ tuples in~$S$ with score at least~$\tau \doteq \AggregationFunt(\underline{y}_1, \ldots, \underline{y}_m)$, where the~$\underline{y}_j$ is the score of the last item in~$\DataList{j}$ retrieved under sorted access.

\vspace{-2mm}
\begin{algorithm}[!ht]
    \caption{Threshold Algorithm~$\algoTA$~\citep{FaginLN03} }
    \label{algo: threshold algorithm}
    \begin{algorithmic}[1]
        \STATE {\bfseries Input:} Sorted array~$\DataList{j}$ and inverted index~$\invertedIndex{j}$, $\forall j \in [t]$.
        \vspace{-3mm}
        \STATE $S \leftarrow \varnothing$. 
        \REPEAT
            \FOR{each $j \in [t]$} \label{line: threshold algorithm round start}
                \STATE Retrieve a tuple from~$\DataList{j}$ via sorted access, and denote the returned tuple as~$(i, \val)$.
                
                \STATE Retrieve from~$\DataList{1}, \ldots, \DataList{t}$ by random access (with the help of~$\invertedIndex{1}, \ldots, \invertedIndex{t}$) all attributes of item~$i$, to compute~$\AggregationFunt(\vec{y}_{i})$.

                \STATE If~$\AggregationFunt(\vec{y}_{i})$ is among the-$k$ highest scores seen so far, add~$\paren{i, \AggregationFunt(\vec{y}_{i})}$ to~$S$; if $|S| > k$, remove the tuple with lowest score from~$S$ . %
            \ENDFOR
            \STATE 
            \label{line: threshold}
            For each~$\DataList{j}$, let~$\underline{y}_j \doteq \vec{y}_{i}[j]$, where~$i$ is the last item seen in~$\DataList{j}$ under sorted access.
            \STATE 
            \label{line: threshold algorithm round end}
            Define the threshold~$\tau \doteq \AggregationFunt(\underline{y}_1, \ldots, \underline{y}_m)$.
        \UNTIL{there are~$k$ tuples in~$S$ with score at least~$\tau$.}
        \STATE Return the set of items contained in the tuples in~$S$.
    \end{algorithmic}
\end{algorithm}
\vspace{-2mm}

The correctness of the algorithm is obvious: when the algorithm stops, since~$\AggregationFunt$ is monotone, the scores of all unseen items are at most~$\tau$, which are lower than the scores of all tuples in~$S$.

\vspace{-2mm}
\paragraph{Access Cost.}
\citeauthor{FaginLN03}~\citeyearpar{FaginLN03} did not provide asymptotic bound for the access cost.
Instead, they proved that~$\algoTA$ is \emph{instance optimal}.
Informally, instance optimally implies that for every algorithm~$\cA$ which solves the top-$k$ selection problem correctly and whose first access to an item must be sorted access as opposed to random access, 
the access cost of~$\algoTA$ is at most the access cost of~$\cA$ (up to some multiplicative constant). 
In Section~\ref{sec: solution}, we apply a different technique to asymptotically bound the access cost of our algorithm.

\subsection{\bf One-shot Private Top-$k$ Algorithm}

We review an existing framework for the differentially private top-$k$ selection algorithms~\citep{DurfeeR19, QiaoSZ21}. 
The framework, described in Algorithm~\ref{algo: top k algorithm},
does not consider a specific data access model, and instead needs to learn all entries of~$\hist$.

\begin{algorithm}[!ht]
    \caption{Private Top-$k$ Algorithm~$\cM$}
    \label{algo: top k algorithm}
    \begin{algorithmic}[1]
        \STATE {\bfseries Input:} vector~$\hist$
        \FOR{each item~$i \in [m]$}   
            \STATE~$\vec{v}[i] \leftarrow \hist[i] + Z_i$, where~$Z_i$ is an independent noise random variable; 
        \ENDFOR
        \STATE Return a set~$S$ of~$k$ items that the maximizes the~$\vec{v}[i]$'s.
    \end{algorithmic}
\end{algorithm}

\begin{definition}[Noise Distributions]
    Given parameter~$b \in \R$, the Laplace distribution,~$\LapNoise{b}$, and the Gumbel distribution,~$\GumbelNoise{b}$,
     have probability density functions~$
        p(z) = \frac{1}{2b} \cdot \exp \PAREN{ - \frac{|z|}{b} },
    $ 
    $\forall z \in \R,$ and~$
        p(z) = \frac{1}{b} \cdot \exp \PAREN{ - \PAREN{ \frac{z}{b} + \exp \PAREN{ - \frac{z}{b} } }  },
    $ 
    $\forall z \in \R,$ respectively.
\end{definition}

Candidates noise distributions for~$Z_i$ in Algorithm~\ref{algo: top k algorithm} include~$\LapNoise{1 / \eps}$~\citep{QiaoSZ21} and~$\GumbelNoise{1 / \eps}$~\citep{DurfeeR19}.
The corresponding privacy guarantees, are stated as follows.

\vspace{2mm}
\begin{fact}[\citep{QiaoSZ21}]
    Assume that~$Z_i \sim \LapNoise{1 / \eps}$, then~$\cM$ is~$2 k \eps$-DP.
    Given~$\delta \in [0, 0.05]$, if it holds that~$m \ge 2$ and~$8 \eps \sqrt{k \log \paren{ m / \delta}} \le 0.2$, then~$\cM$ also satisfies~$\PAREN{8 \eps \sqrt{k \log \paren{ m / \delta}}, \delta}$-DP.
\end{fact}

\begin{fact}[\citep{DurfeeR19}]
    Assume that~$Z_i \sim \GumbelNoise{1 / \eps}$. 
    For each~$\delta \in [0, 1]$,~$\cM$ is~$\paren{\eps'', \delta}$-DP, where~$\eps'' \doteq \min \set{ k \eps, k \eps \PAREN{\frac{e^\eps - 1}{e^\eps + 1}} + \eps \sqrt{k \ln \frac{1}{\delta} } }$. 
\end{fact}

Next, we discuss their utility guarantees. 

\begin{fact}
    \label{fact: report noisy max k utility}
    Given~$\beta \in (0, 1)$, if the~$Z_i \sim \LapNoise{1 / \eps}$, or~$\GumbelNoise{1 / \eps}$, then with probability at least~$1 - \beta$, the returned solution by Algorithm~\ref{algo: top k algorithm} is~$\paren{\alpha, k}$-accurate, for
    $
        \alpha \in O \PAREN{ \frac{ \ln \paren{m / \beta} }{\eps } }. 
    $
\end{fact}

\begin{remark}
    Compared to Laplace noise, the Gumbel noise allows the algorithm to return a ranked list of indices, instead of a set which contains on order information.
    For consistency of presentation, we assume that Algorithm~\ref{algo: top k algorithm} returns a set for both choices. 
\end{remark}

%% file: top-k-algorithm.tex
\vspace{-2mm}
\section{Sublinear Access and Time Algorithm}
\label{sec: solution}

In this section, we present an algorithm for top-$k$ selection problem, which achieves optimal privacy-utility trade-offs, and with high probability, has an expected access cost~$O \paren{ \sqrt{mk} }$ and computation time~${O} \paren{ \sqrt{mk} \log \log m }$.
Our presentation follows two steps: we first present a strawman  algorithm with sublinear access cost but only linear computation; next we show how to improve its time complexity to~${O} \paren{ \sqrt{mk} \log \log m }$.

\vspace{-2mm}
\subsection{A Strawman Approach}
\label{subsec: starwman solution}

A natural idea is to combine the threshold algorithm~$\algoTA$ with the oneshot private top-$k$ algorithm.
Each item~$i \in [m]$ now has two attributes, namely,~$\hist[i]$ and~$Z_{i}$, where~$Z_{i}$ is sampled independently from~$\LapNoise{1 / \eps}$ or~$\GumbelNoise{1 / \eps}$.

Since the histogram $\hist$ is stored in a database management system, which allows for two types of access: sorted access and random access, we can think of this as a sorted array~$\DataList{1}$ of~$m$ tuples, each of the form~$(i, \val) \in [m] \times \R$, where~$\val$ equals~$\hist[i]$.%
The tuples in~$\DataList{1}$ are sorted in descending order by their~$\val$'s.
Further,~$\DataList{1}$ has an inverted index~$\invertedIndex{1}$ to support random access.

Additionally, we can construct another sorted array~$\DataList{2}$ of~$m$ tuples, each of the form~$(i, \val) \in [m] \times \R$, where~$\val$ equals~$Z_{i}$.
The tuples in~$\DataList{2}$ are also sorted in descending order by their~$\val$'s.
$\DataList{2}$ also has an inverted index~$\invertedIndex{2}$ to support random access.

Then we can run the algorithm~$\algoTA$, with input~$\DataList{1}, \DataList{2}$, $\invertedIndex{1}, \invertedIndex{2}$, and an aggregating function 
$
    \AggregationFunt(\hist[i],Z_{i})
    \doteq \hist[i] + Z_{i}.
$
It is easy to see that~$\AggregationFunt$ is monotone.
The pseudo-code is in Algorithm~\ref{algo: modified threshold algorithm}. 

\begin{algorithm}[!ht]
    \caption{Private Threshold Algorithm~$\algoPrivTA$}
    \label{algo: modified threshold algorithm}
    \begin{algorithmic}[1]
        \STATE Let~$I_1 = 1, I_2 = 2, \ldots, I_m = m$. 
        Generate~$m$ tuples~$(I_1, Z_1), \ldots, (I_m, Z_m)$, where the~$Z_i$'s are i.i.d.~random variables;
        sort the tuples in descending order by the values of the~$Z_i$'s, denote the sorted sequence by~$\PAREN{I_{(1)}, Z_{(1)}}, \ldots, \PAREN{I_{(m)}, Z_{(m)}}$, and store this sequence in an array~$\DataList{2}$;
        construct~$\invertedIndex{2} : [m] \rightarrow [m]$, s.t.,~$\DataList{2}[\invertedIndex{2}(i)] = (i, Z_i)$ for each item~$i \in [m]$.
        
        \STATE Run Algorithm~\ref{algo: threshold algorithm} on input~$\DataList{1}, \DataList{2}, \invertedIndex{1}, \invertedIndex{2}$, with an aggregation function~$\AggregationFunt \big( \paren{\hist[i], Z_{i}} \big) \doteq \hist[i] + Z_{i}$;
    \end{algorithmic}
\end{algorithm}

\vspace{-3mm}
\paragraph{Privacy and Utility Guarantee.}
The privacy and utility guarantee of the algorithm inherits directly from Algorithm~\ref{algo: top k algorithm}. 

\vspace{-3mm}
\paragraph{Access and Time Complexity.} 
It is easier to first discuss the time complexity and then the access cost.
Generating the random variables takes~$O(m)$ time, and sorting them takes~$O(m \log m)$ time.
Hence the total running time is bounded by~$O(m \log m)$. 

It remains to discuss the number of accesses the algorithm performs on~$\DataList{1}$.
Our analysis relies on the following important observation.

\begin{lemma}
    \label{lemma: permutaion and independence}
    The~$I_{(1)}, \ldots, I_{(m)}$ are distributed uniformly over all possible permutations over~$[m]$, and are independent of the random variables~$Z_{(1)}, \ldots, Z_{(m)}$. 
\end{lemma}

Intuitively, the claim holds since each~$Z_i$ in Algorithm~\ref{algo: modified threshold algorithm} follows the same distribution independently.
The proof of the lemma is included in Appendix~\ref{appendix: proof for subsec: starwman solution}.

\begin{theorem}
    \label{theorem: access cost of algo: modified threshold algorithm}
    The expected access cost of Algorithm~\ref{algo: modified threshold algorithm} on~$\DataList{1}$,~$\E{\AccessCost{\algoPrivTA}{\DataList{1}}}$, is bounded by~$O \PAREN{ \sqrt{m k}}$.
\end{theorem}

The rigorous proof of the Theorem is presented in Appendix~\ref{appendix: proof for subsec: starwman solution}. 
    Here we offer an intuitive and informal explanation. 
    Let~$\cS_r \doteq \set{ i_{(1)}, \ldots, i_{(r)} }$ be the top-$r$ items with highest scores in~$\hist$, and~$I_{(1:r)} = \set{ I_{(1)}, \ldots, I_{(r)} }$ be the items in the first~$r$ tuples in the array~$\DataList{2}$.
    Since~$I_{(1:r)}$ is a uniform random subset of~$[m]$, $\cS_r \cap I_{(1:r)}$ has expected size~$r \cdot \PAREN{r / m} = \PAREN{r^2 / m}$, which equals~$k$ when~$r = \sqrt{mk}$.
    Applying a technique introduced by~\citet{Fagin99}, we can show that when $\card{\cS_r \cap I_{(1:r)}} \ge k$, the algorithm will not access any item outside~$\cS_r \cup I_{(1:r)}$, since any such item will have a score lower than or equal to any item in~$\cS_r \cap I_{(1:r)}$.
    Therefore the algorithm should have access cost roughly~$O \PAREN{ \sqrt{m k}}$.

\paragraph{Application.}

    We discuss an interesting application of our algorithm to the exponential mechanism~\citep{McSherryT07}, a fundamental technique in differential privacy to choose a single item from a set of items.

    Following the setup in this paper, the exponential mechanism works as follows: it selects an item~$i \in [m]$ with probability proportional to~$e^{ \eps \cdot \hist[i] }$.
    Moreover, \citet{DurfeeR19} show that the exponential mechanism is equivalent to Algorithm~\ref{algo: top k algorithm} with~$k = 1$ and~$Z_i \sim \GumbelNoise{1 / \eps}, \forall i \in [m]$. 
    This variant of Algorithm~\ref{algo: top k algorithm} is commonly referred to as the Report-Noisy-Max algorithm with Gumbel noise.
    Applying the same 
    $k$ and the~$Z_i$'s to Algorithm~\ref{algo: modified threshold algorithm}, Theorem~\ref{theorem: access cost of algo: modified threshold algorithm} immediately implies the following corollary.

\begin{corollary}
    \label{corollary: access of exponential mechanism}
        When given access to both sorted and random access to data, the exponential mechanism has expected access cost~$O(\sqrt{m})$.
\end{corollary}

\subsection{An Online Sampling Approach}
\label{subsec: An Online Sampling Approach}

Pre-generating all $m$ noise values may be excessive. 
For problems with small values of~$k$, e.g.,~$k = 10$, the Algorithm~\ref{algo: modified threshold algorithm} may need to know only a small subset of tuples in~$\DataList{2}$.
It is of interest whether we can also reduce the expected number of noisy random variables generated to~${O} \paren{ \sqrt{mk} }$, by constructing the~$\DataList{2}$ (and~$\invertedIndex{2}$) on the fly.

    One can consider applying existing algorithms (such as those presented in~\citealt{Lurie1972MachineGenerationOO, Devroye86}) to generate the random variables~$\PAREN{I_{(1)}, Z_{(1)}}, \ldots, \PAREN{I_{(m)}, Z_{(m)}}$ sequentially, one tuple at a time, each taking~$O(1)$ time. 
    However, since the threshold algorithm relies on non-sequential accesses to the variables (due to the \emph{random access} operation), these algorithms cannot be applied to reduce the number of variables generated to~$O \PAREN{ \sqrt{mk} }$.
    In this section, instead, we present an algorithm that can generate the variables in an arbitrary order and ``on the fly''.
The main result of this section is stated as follows.

\begin{theorem}
    \label{theorem: oracle}
    There is an algorithm~$\algoOracle$, that, %
    \begin{itemize}[leftmargin=4.5mm, topsep=2pt, itemsep=2pt, partopsep=2pt, parsep=2pt]

        \item does not require to pre-generate~$\DataList{2}$;
        
        \item answers sorted access and random access query to~$\DataList{2}$ in~$O(\log \log m)$ time in expectation. 
        
    \end{itemize}
    Further, the tuples returned by~$\algoOracle$ have the same marginal distribution as those generated by Algorithm~\ref{algo: modified threshold algorithm}.
\end{theorem}

There are two key ingredients for constructing~$\algoOracle$.

\paragraph{Sampling the \texorpdfstring{$I_{(j)}$}{Lg}'s.}

The first ingredient is Lemma~\ref{lemma: permutaion and independence}, which allows~$\algoOracle$ to sample the~$I_{(j)}$ and the~$Z_{(j)}$ independently according to their marginal distributions, without changing the joint distribution of the~$I_{(j)}$ and the~$Z_{(j)}$.
The lemma states that the~$I_{(j)}$'s are distributed uniformly over all possible permutations over~$[m]$.
It is not hard to sample an~$I_{(j)}$ on the fly: let~$\cJ$ be the set of indexes such that the values of~$I_{(j')}, j' \in \cJ$ have been determined; if~$j \notin \cJ$, then $I_{(j)}$ just distributes uniformly over the subset of unseen items, i.e.,~$[m] \setminus I_{(\cJ)}$, where~$I_{(\cJ)} \doteq \set{I_{(j')}:  j' \in \cJ}$. 
Correspondingly, we can also construct the inverted index~$\invertedIndex{2}$ on the fly: given an item~$i \in [m]$, if it has not been encountered, then~$\invertedIndex{2}(i)$ should equal one of the undetermined indexes, namely~$[m] \setminus \cJ$, uniformly at random.

\paragraph{Sampling the \texorpdfstring{$Z_{(j)}$}{Lg}'s.}

The second ingredient is an algorithm~$\algoOrderStat$ which generates the~$Z_{(j)}$'s on the fly.%
Formally, for each~$\cJ \subseteq [m]$, define~$Z_{(\cJ)} \doteq \paren{ Z_{(j)}, j \in \cJ}$, and let~$z_{(\cJ)}$ refer to a vector~$\paren{ z_{(j)}, j \in \cJ} \in \R^{|\cJ|}$. 
Denote by~$p_{ {\textstyle\mathstrut} Z_{(\cJ)}} (\cdot)$ the marginal density of~$Z_{(\cJ)}$, induced by the generating procedure of Algorithm~\ref{algo: modified threshold algorithm}. 
Call~$z_{(\cJ)}$ a feasible realization of~$Z_{(\cJ)}$, if~$p_{ {\textstyle\mathstrut} Z_{(\cJ)}} ( z_{(\cJ)} ) > 0$.
Given such a feasible realization, let $p_{ {\textstyle\mathstrut} Z_{(j)} \mid  Z_{(\cJ)}} \paren{ z_{(j)} \mid z_{(\cJ)} }$ be the density function of~$Z_{(j)}$, conditioned on~$Z_{(\cJ)} = z_{(\cJ)}$. 
The property of~$\algoOrderStat$ is stated as follows. 

\begin{lemma}
    \label{lemma: sampling zj}
    For each $\cJ \subseteq [m]$ s.t., $\cJ \neq [m]$, each~$j \in [m] \setminus \cJ$, and each feasible realization $z_{(\cJ)}$ of $Z_{(\cJ)}$, 
    $\algoOrderStat$ samples a random variable with the conditional density 
    $
        p_{ {\textstyle\mathstrut} Z_{(j)} \mid  Z_{(\cJ)}} \paren{ z_{(j)} \mid z_{(\cJ)} }
    $
    in~$O(\log \log m)$ expected time.
\end{lemma}

The proof of the Lemma is discussed in Section~\ref{subsec: sampling Zj} .
Now, we return to the construction of~$\algoOracle$.
The algorithm is described in~Algorithm~\ref{algo: online sampling algorithm}.

\begin{algorithm}[!ht]
    \caption{Algorithm~$\algoOracle$}
    \label{algo: online sampling algorithm}
    {\it Initialization}
    \begin{algorithmic}[1]
        \STATE~$\cJ \leftarrow \varnothing$,~$\idx \leftarrow 0$; 
        \STATE~$\DataList{2}[i] \leftarrow nil, \invertedIndex{2}(i) \leftarrow nil,\, \forall i \in [m]$
    \end{algorithmic}
    {\it Sorted Access}
    \begin{algorithmic}[1]
        \STATE~$\idx \leftarrow \idx + 1$;
        \IF{$\idx \notin \cJ$} 
            \STATE Sample~$I_{(\idx)}$ uniformly from~$[m] \setminus I_{(\cJ)}$; 
            \STATE Invoke~$\algoOrderStat$ to sample~$Z_{(\idx)}$;
            \STATE $\invertedIndex{2}(I_{(\idx)}) \leftarrow \idx$;
            \STATE $\DataList{2}[\idx] \leftarrow \paren{I_{(\idx)}, Z_{(\idx)}}$;~$\cJ \leftarrow \cJ \cup \set{ \idx }$.
        \ENDIF
        \STATE {\bf return}~$\DataList{2}[\idx]$ 
    \end{algorithmic}
    \mbox{\it Random Access (Input: item $i \in [m]$)}
    \vspace{-3.5mm}
    \begin{algorithmic}[1]
        \IF{$\invertedIndex{2}(i) = nil$} 
            \STATE Sample~$j$ uniformly from~$[m] \setminus \cJ$; 
            \STATE Invoke~$\algoOrderStat$ to sample~$Z_{(j)}$;
            \STATE $\invertedIndex{2}(i) \leftarrow j$;
            \STATE $\DataList{2}[j] \leftarrow \paren{i, Z_{(j)}}$;~$\cJ \leftarrow \cJ \cup \set{ j }$.
        \ENDIF
        \STATE {\bf return}~$\DataList{2}[\invertedIndex{2}(i)]$ 
    \end{algorithmic}
\end{algorithm}
\vspace{-2mm}

\vspace{-1mm}
\paragraph{Initialization.} 
The algorithm creates an empty array~$\DataList{2}$ and an empty inverted index~$\invertedIndex{2}$.
Further, it creates a set~$\cJ$, to record the positions of~$\DataList{2}$ which are already sampled, and a variable~$\idx$, to record the last visited position by sorted access. 
In practice,~$\DataList{2}$ and~$\invertedIndex{2}$ need not to be physically initialized, and can be implemented by hash sets with constant initialization time.

\vspace{-2mm}
\paragraph{Sorted Access.} 
Indeed, it is trivial to handle the sorted access.
We just maintain an index,~$\idx \in \N$, of last tuple returned by sorted access.
When a new request of sorted access arrives, we increase~$\idx$ by~$1$. 
If~$\idx \in \cJ$, then~$\algoOracle$ returns~$\DataList{2}[\idx]$ directly; otherwise, it generates~$\DataList{2}[\idx]$ before returning it. 

\vspace{-2mm}
\paragraph{Random Access.}
A random access request comes with a reference to an item~$i \in [m]$.
We need to identify the tuple~$\DataList{2}[j] = \paren{I_{(j)}, Z_{(j)}}$ s.t.,~$I_{(j)} = i$. 
There are two cases: if item~$i$ has been encountered previously ($\invertedIndex{2}(i) \neq nil$), then~$\algoOracle$ returns~$\DataList{2}[\invertedIndex{2}(i)]$ directly; otherwise, ~$\algoOracle$ randomly pick an index~$j$ from~$[m] \setminus \cJ$, and set~$I_{(j)} \leftarrow i$,~$\invertedIndex{2}(i) \leftarrow j$, and  calls~$\algoOrderStat$ to generate~$Z_{(j)}$.
Finally, it returns~$\DataList{2}[j]$.

\vspace{-1mm}
\subsubsection{Sampling Ordered Noises}
\label{subsec: sampling Zj}

In this section, we show how to construct $\algoOrderStat$ and prove Lemma~\ref{lemma: sampling zj}.
Deciding the conditional distribution of the~$Z_{(j)}$'s and sampling them directly from such distribution can be non-trivial.
As a result, we follow the three-step approach outlined below: 

\begin{itemize}[leftmargin=4.5mm, topsep=2pt, itemsep=2pt, partopsep=2pt, parsep=2pt]
    \item 
    {\it Transform~$U_{(j)}$ to~$Z_{(j)}$}:
    we show 
    that the sorted sequence~$Z_{(1)}, \ldots, Z_{(m)}$ 
    can be transformed from a sequence~$U_{(1)}, \ldots, U_{(m)}$ of sorted
    independent uniform random variables.

    \item 
    {\it Distribution of~$U_{(j)}$}:
    to avoid generating the entire sequence of random variables, we study the distribution of~$U_{(j)}$, conditioned on a set of~$U_{(j')}$ which have already been sampled. 
    
    \item 
    {\it Sampling~$U_{(j)}$}:
    we show how to sample an~$U_{(j)}$ from such a distribution in~$O(\log \log m)$ expected time.
\end{itemize}

\paragraph{Transform~$U_{(j)}$ to~$Z_{(j)}$.}
Since all potential noise distributions ($\LapNoise{1 / \eps}$ or~$\GumbelNoise{1 / \eps}$) of the~$Z_i$'s (the noise random variables, before sorting) have continuous cumulative distribution function, 
we can sample them indirectly via uniform random variables, based on the \emph{inversion method}. 

\begin{fact}[Inversion Method~\citep{Devroye86}]
    \label{fact: random varaible generation by uniform}
    Let~$F$ be a continuous cumulative distribution function on~$\R$ with inverse~$F^{-1}$ defined by
    \vspace{-1mm}
    \begin{align*}
        F^{-1}(u) \doteq \inf \set{ x : F (x) = u, 0 < u <1 }. 
        \vspace{-1mm}
    \end{align*}
    If~$U$ is a uniform $[0,1]$ random variable, then $F^{-1}(U)$ has distribution function~$F$.
\end{fact}

    As a result, 
    an ordered sequence of random variables can also be generated via the inversion method. 
    \begin{fact}[\citealt{gerontidis1982monte}]
        \label{fact: inversion method for order statistics}
        Let~$U_1, \ldots, U_m$ be independent uniform random variables on~$[0, 1]$, and~$U_{(1)}, \ldots, U_{(m)}$ the corresponding sorted sequence in descending order.
        Let~$F$ be a continuous cumulative distribution function shared by the random variables~$Z_1, \ldots, Z_m$.
        Then the sequence~$F^{-1}(U_{(1)}), \ldots, F^{-1}(U_{(m)})$ has the same distribution as the sequence~$Z_{(1)}, \ldots, Z_{(m)}$. 
    \end{fact}
    For completeness, we include a proof for this fact in Appendix~\ref{appendix: proof for subsec: An Online Sampling Approach}. 
    We can thus sample~$U_{(j)}$ first and then compute~$F^{-1}(U_{(j)})$ to generate~$Z_{(j)}$.
    It remains to study the distribution of~$U_{(j)}$, 
    and the algorithm for sampling a random variable efficiently from such distribution.

\paragraph{Distribution of~$U_{(j)}$.}
Recall that~$\cJ$ is the set of indexes which have been previously queried. 
Denote $U_{(\cJ)} \doteq$ $\set{U_{(j')} : j' \in \cJ}$ a shorthand for the order random variables that have been sampled.
Further, write~$u_{(\cJ)} \doteq \set{u_{(j')} : j' \in \cJ} \in [0, 1]^{\card{\cJ}}$ as a set of numbers within~$[0, 1]$, indexed by~$\cJ$.
Call~$u_{(\cJ)}$ a \emph{feasible realization} of~$U_{(\cJ)}$, if for each~$j, j' \in \cJ$ s.t.~$j < j'$, it holds that~$u_{(j)} \ge u_{(j')}$.

Given a new query index~$j \in [m] \setminus \cJ$, we are interested in the conditional probability density,~$p_{ {\textstyle\mathstrut} U_{(j)} \mid  U_{(\cJ)}} \paren{ u_{(j)} \mid u_{(\cJ)} }$, of~$U_{(j)}$, given the occurrence of a feasible realization~$u_{(\cJ)}$ of~$U_{(\cJ)}$.
For ease of reading, we omit the subscripts of the conditional probability densities, whenever their meaning can be unambiguously determined from their parameters.

Depending on the relative position of~$j$ w.r.t.~the indexes in~$\cJ$, we consider the following three cases: 
\begin{itemize}[leftmargin=4.5mm, topsep=1pt, itemsep=2pt, partopsep=2pt, parsep=2pt]
    \item $\cJ$ is empty.
    It reduces to study the un-conditional probability density~$p( u_{(j)} )$ of~$U_{(j)}$. 
        
    \item $\cJ$ is not empty, and~$j$ is greater than the largest index in~$\cJ$; in this case, $j$ has a predecessor (the largest index that is smaller than~$j$), denoted by~$\ell$, in~$\cJ$.
    
    \item $\cJ$ is not empty, and~$j$ is smaller than the largest index in~$\cJ$; in this case, $j$ has both a predecessor, denoted by~$\ell$, and a successor (the smallest index that is larger than~$j$), denoted by~$r$, in~$\cJ$.
\end{itemize}

Hereafter, if~$\cJ \neq \varnothing$, we consider only feasible realization~$u_{(\cJ)}$ of~$U_{(\cJ)}$. 
    The probability densities corresponding to these three cases are given thus.

\begin{theorem}
    \label{theorem: conditional distribution of the order statistics}
    \begin{enumerate}[wide, labelwidth=0pt, labelindent=0pt, label=\normalfont{(\arabic*)}]
        \item If~$\cJ$ is empty, then the density~$p(u_{(j)})$ of~$U_{(j)}$ is given by:~$\forall\, u_{(j)} \in [0, 1]$, 
            \begin{align}
                \label{equa: density of jth of uniform random varaibles}
                \begin{array}{l}
                     p(u_{(j)}) = \frac{m!}{\paren{j - 1}! \paren{m - j}!}\cdot \paren{1 - u_{(j)} }^{j - 1} \paren{u_{(j)}}^{m - j}. 
                \end{array}
            \end{align}
            
        \vspace{-2mm}
        \item  If $\cJ$ is not empty, and~$j$ is greater than the largest index in~$\cJ$, then given~$U_{(\ell)} = u_{(\ell)}$,~$U_{(j)}$ is independent of all other random variables~$U_{(j')}$ for all~$j' \in \cJ \setminus \set{ \ell }$, i.e.,~$p \PAREN{ u_{(j)} \mid u_{(\cJ)} } = p \PAREN{ u_{(j)} \mid u_{(\ell)} }$; further, for each~$u_{(j)} \in [ 0, u_{(\ell)} ]$, 
            \begin{align}
                \label{equa: conditional density case two}
                \begin{array}{l}
                        p \PAREN{ u_{(j)} \mid u_{(\ell)} } = 
                        \frac{(m - \ell)!}{ \paren{ j - \ell  - 1}! \paren{m - j}!} \cdot 
                    \\ \qquad \qquad \qquad
                        \PAREN{
                            \frac{
                                u_{(\ell)} - u_{(j)}
                            }{
                                u_{(\ell)} 
                            }
                        }^{j - \ell  - 1} 
                        \PAREN{
                            \frac{
                                u_{(j)} 
                            }{
                                u_{(\ell)} 
                            }
                        }^{m - j} 
                        \frac{1}{u_{(\ell)}}. 
                \end{array}
            \end{align}

        \vspace{-2mm}  
        \item If $\cJ$ is not empty, and~$j$ is smaller than the largest index in~$\cJ$, then given~$U_{(\ell)} = u_{(\ell)}$ and~$U_{(r)} = u_{(r)}$,~$U_{(j)}$ is independent of all other random variables  ~$U_{(j')}$ for all~$j' \in \cJ \setminus \set{ \ell,  r}$, i.e.,~$p \PAREN{ u_{(j)} \mid u_{(\cJ)} } = p \PAREN{ u_{(j)} \mid u_{(\ell)}, u_{(r)} }$; further, for each~$u_{(j)} \in [ u_{(r)}, u_{(\ell)} ]$, 
            \begin{align}
                \label{equa: conditional density case three}
                \hspace{-5mm}
                \begin{array}{lr}
                    p \PAREN{ u_{(j)} \mid u_{(\ell)}, u_{(r)} } 
                    = 
                    \frac{
                        \paren{r - \ell - 1}!
                    }{
                        \paren{j - \ell - 1}! \paren{r - j - 1}!
                    } \cdot \\
                    \qquad \quad
                    \PAREN{\frac{u_{(\ell)} - u_{(j)}}{u_{(\ell)} - u_{(r)}}}^{j - \ell - 1} 
                    \PAREN{\frac{u_{(j)} - u_{(r)}}{u_{(\ell)} - u_{(r)}}}^{r - j - 1}
                    \frac{1}{u_{(\ell)} - u_{(r)}}.
                \end{array}
            \end{align}

    \end{enumerate}
 
\end{theorem}

The theorem removes the dependency of~$U_{(j)}$ from all but at most two variables in~$U_{(\cJ)}$.
The detailed proof is non-trivial and can be found in  Appendix~\ref{appendix: proof for theorem: conditional distribution of the order statistics}. 
Assuming that~$U_{(j)}$ depends on at most two variables in~$U_{(\cJ)}$, 
we can provide an informal, but intuitive, explanation of the conditional probability densities. 
Take the Equation~(\ref{equa: conditional density case three}) for example.
Conditioned on~$U_{(r)} = u_{(r)}$ and~$U_{(\ell)} = u_{(\ell)}$, $r - \ell - 1$ 
uniform random variables
fall into the interval~$[u_{(r)}, u_{(\ell)}]$.
Of these~$r - \ell - 1$ random variables, $j - \ell - 1$ of them are $\ge u_i$, and~$r - j - 1$ of them are~$< u_i$.
The number of possible combinations is given by
$\frac{
    \paren{r - \ell - 1}!
}{
    \paren{j - \ell - 1}! \paren{r - j - 1}!
}.$
For a fixed combination, the former happens with probability~$\PAREN{\frac{u_{(\ell)} - u_{(j)}}{u_{(\ell)} - u_{(r)}}}^{j - \ell - 1}$, the latter happens with probability~$\PAREN{\frac{u_{(j)} - u_{(r)}}{u_{(\ell)} - u_{(r)}}}^{r - j - 1}$, and the probability density of~$U_{(j)} = u_{(j)}$ is~$\frac{1}{u_{(\ell)} - u_{(r)}}$. 

\paragraph{Sampling~$U_{(j)}$.}
We now discuss how to sample the~$U_{(j)}$ efficiently from their conditional distributions.
First, note that determining the conditional distributions may need to find the index~$j$'s predecessor or successor in~$\cJ$.
This can be done by Van Emde Boas tree~\citep{Boas75} in~$O(\log \log m)$ time.
Next, we show that sampling from such conditional distributions takes~$O(1)$ expected time. 
Specifically, we will sample random variables with \emph{Beta distributions}, then convert them into ones which follow desired conditional distributions. 

\begin{definition}[Beta Distribution~\citep{Ross2018}]
    The beta distribution~$\BetaDist{\alpha}{\beta}$ is a distribution defined on~$[0, 1]$ whose density is given by 
    \vspace{-1mm}
    \begin{align}
        \begin{array}{c}
            p(x) = \frac{
                x^{\alpha - 1} \paren{1 - x}^{\beta - 1}
            }{
                \BetaFunt{\alpha}{\beta}
            }, \qquad \forall x \in [0, 1], 
        \end{array}
    \vspace{-1mm}
    \end{align}
    where~$\alpha, \beta > 0$ are \emph{shape parameters},~$\BetaFunt{\alpha}{\beta} \doteq \int_0^1 x^{\alpha - 1} \paren{1 - x}^{\beta - 1} \, dx$ is a normalisation constant. 
\end{definition}

It is known that, when~$\alpha \ge 1, \beta \ge 1$, a random variable~$X \sim \BetaDist{\alpha}{\beta}$ can be generated in~$O(1)$ expected time~\citep{Devroye86, Gentle2009}.

\begin{theorem}
    \label{theorem: converting beta distribitons}
    Assume that~$\ell < j < r \le m$, and~$1 \ge u_{(\ell)} > u_{(r)} \ge 0$.
    Then
    \begin{enumerate}[leftmargin=4.5mm, topsep=1pt, itemsep=2pt, partopsep=2pt, parsep=1pt]
        \vspace{-1mm}
        \item If~$X \sim \BetaDist{m - j + 1}{j}$, then the density function of~$X$ is the same as Equation~(\ref{equa: density of jth of uniform random varaibles}). 
        
        \vspace{-1mm}
        \item If~$X \sim \BetaDist{m - j + 1}{j - \ell}$, then the density function of~$Y \doteq u_{(\ell)} \cdot X $ is the same as Equation~(\ref{equa: conditional density case two}). 
        
        \vspace{-1mm}
        \item If~$X \sim \BetaDist{r - j}{j - \ell}$, then the density function of~$Y \doteq \paren{u_{(\ell)} - u_{(r)}} \cdot X$ is the same as Equation~(\ref{equa: conditional density case three}). 
    \end{enumerate}
\end{theorem}

The proof of the Theorem is included in Appendix~\ref{appendix: proof for subsec: An Online Sampling Approach}.

%% file: lower-bounds.tex
\section{Lower Bounds} \label{sec: lower bounds}

In this section, we generate the lower bounds for the problem. 
Following the setting in Section~\ref{sec: problem definition}, since~$\hist$ is the sum of voting vectors of~$n$ clients, we have~$|| \hist ||_\infty \le n$.
It follows that each $S \in \binom{[m]}{k}$ is $\paren{n, k}$-accurate.
All lower bounds in this section hold for algorithms that are~$\paren{n - O(1), k}$-accurate, which is just slightly better than the trivial error guarantee.

\vspace{-2mm}
\subsection{Random Access}
\label{subsec: random access}

We first present a lower bound for the random access case. 

\begin{theorem}
    \label{theorem: lower bound for random access}
    Assume that~$0 \le \beta < 0.1$. 
    Let~$\cA$ be an algorithm that has only~\emph{random access} to~$\hist$, does not return items which it has not seen , and for each input, with probability at least~$1 - \beta$, returns a solution that is~$\paren{n - 1, k}$-accurate.
    Then there exists a family of histograms~$\cH$, and a distribution~$\mu$ on~$\cH$, if~$\hist$ is sampled from~$\cH$ according to distribution~$\mu$, it holds that
    \begin{align} \small
        \label{ineq: lower bound on the random access cost}
            \E{\hist}{ cost \PAREN{\cA, \hist} } 
            \in 
            \Omega \PAREN{ m }. 
    \end{align}
\end{theorem}

Note that the theorem does not even require~$\cA$ to be a~$(\eps, \delta)$-DP algorithm.
The proof of the theorem is in Appendix~\ref{appendix: proof for subsec: random access}.
At a high level, our construction focuses on a family of histograms the values of whose entries are either~$n$ or~$0$.
Further, if an~$\hist$ is sampled from~$\cH$, there are roughly~$2k$ entries of~$\hist$ that have value~$n$, and those entries appear at random positions of~$\hist$, so that~$\cA$ is unlikely to identify more than~$k$ of them, before learning the values of~$\Omega(m)$ entries.

\vspace{-2mm}
\subsection{Sorted Access} 
\label{subsec: sorted access}

The lower bound for the sorted access case relies on the following lemma. 

\begin{lemma}
    \label{lemma: probability of reporting back candiates}
    Let~$\cA$ be an~$\paren{\eps, \delta}$-DP algorithm, which for each input histogram, with probability at least~$1 - \beta$, returns a solution that is~$\paren{n - 2, k}$-accurate.
    Let~$S \subseteq [m]$, and~$\hist_{S}$ be a histogram, s.t.
    \vspace{-2mm}
    \begin{equation} \small
        \hist_{S}[i] \doteq 
        \begin{cases}
            n - 1,  & \forall \, i \in S, \\
            0,      & \forall \, i \in \bar{S}, 
        \end{cases}
    \vspace{-1mm}
    \end{equation}
    where~$\bar{S} \doteq [m] \setminus S$.
    Let~$\LowFreqSet$ be a subset of~$S$ sampled uniformly at random from $\binom{S}{|S| / k}$, 
    $\HighFreqSet \doteq \cS \setminus \LowFreqSet$, and~$\hist_{{\HighFreqSet, \LowFreqSet}}$  be a histogram neighboring to~$\hist$ s.t. 
    \vspace{-2mm}
    \begin{equation} \small
        \label{equa: construction of random histogram based on S}
        \hist_{{\HighFreqSet, \LowFreqSet}} [i] \doteq 
        \begin{cases}
            n,              & \forall \, i \in \HighFreqSet, \\
            n - 1,       & \forall \, i \in \LowFreqSet, \\
            0,              & \forall \, i \notin S. 
        \end{cases}
    \vspace{-4mm}
    \end{equation}
    Then, 
    \begin{align} 
        \label{ineq: probability of reporting back candiates}
        \begin{array}{c}
            \P{\LowFreqSet, \cA}{ 
                \cA( \hist_{{\HighFreqSet, \LowFreqSet}} ) 
                \cap \LowFreqSet 
                \neq \emptyset 
            }
        \ge \frac{1 - \beta - \delta - e^{-1}}{e^\eps}\,, 
        \end{array}
    \end{align}
    where the randomness is first over the choice of~$\LowFreqSet$ then over the output of~$\cA$.
\end{lemma}

The formal proof of the lemma is included in Appendix~\ref{appendix: proof for subsec: sorted access}. 
Note that, for each~$i \in \LowFreqSet$, $\hist_{{\HighFreqSet, \LowFreqSet}} [i]$ is among the~$|S| / k$ smallest entries of the~$|S|$ largest entries in~$\hist_{{\HighFreqSet, \LowFreqSet}}$.
The lemma states that the probability that output of~$\cA$ contains some item~$i \in \LowFreqSet$ is not too ``small''. 
    Informally, for each subset~$S_k \in \binom{S}{k}$, when $\LowFreqSet$ is sampled uniformly from~$\binom{S}{|S| / k}$, then the 
    probability that~$S_k \cap \LowFreqSet \neq \emptyset$  is not too ``small'' (observe that~$\E{\card{S_k \cap \LowFreqSet}} = k \cdot \tfrac{|S| / k}{|S|} = 1$). 
    Further, if $\cA$'s output is~$(n - 2, k)$ accurate, then it must belong to~$\binom{S}{k}$.
    So the probability that~$\cA$'s output has non-empty intersection with~$\LowFreqSet$ should not be significantly smaller than the probability that $\cA$'s output is~$(n - 2, k)$ accurate.

\begin{theorem}
    \label{theorem: lower bound for sorted access}
    Let~$\eps, \delta, \beta$ be non-negative parameters, s.t.,~$\eps \in O(1), \delta + \beta \le 0.6$. 
    Let~$\cA$ be an~$\paren{\eps, \delta}$-DP algorithm that has only~\emph{sorted access}, does not return items which it has not seen, and for each input histogram, with probability at least~$1 - \beta$, returns a solution that is~$\paren{n - 2, k}$-accurate.
    Then there exists a family of histograms~$\cH$ so that, if~$\hist$ is sampled uniformly at random from~$\cH$, it holds that
    \begin{align} \small
        \label{ineq: lower bound on the expected sorted access cost}
        \E{\hist}{ cost \PAREN{\cA, \hist} } 
        \in 
        \Omega \PAREN{ m }. 
    \end{align}
\end{theorem}

\begin{proof}
    Let~$S = [m / 2]$, 
    $\LowFreqSet$ be sampled uniformly at random from $\binom{S}{|S| / k}$, 
    and~$\hist_{{\HighFreqSet, \LowFreqSet}}$ be a histogram built as outlined in Equation~(\ref{equa: construction of random histogram based on S}).
    Let~$\cH$ be the collection of all possible outcomes of~$\hist_{{\HighFreqSet, \LowFreqSet}}$.
    Then by Lemma~\ref{lemma: probability of reporting back candiates}, 
    $$ 
    \begin{array}{c}
            \P{\LowFreqSet, \cA}{ 
                \cA( \hist_{{\HighFreqSet, \LowFreqSet}} ) 
                \cap \LowFreqSet 
                \neq \emptyset 
            }
        \ge \frac{1 - \beta - \delta - e^{-1}}{e^\eps}. 
    \end{array}
    $$
    But for each~$i \in \LowFreqSet$,~$\hist_{{\HighFreqSet, \LowFreqSet}}[i]$ is among the top~$\paren{|S| - |S| / k + 1}^{(th)}$ to~${|S|}^{(th)}$ largest numbers in~$\hist_{{\HighFreqSet, \LowFreqSet}}$. 
    Since~$\cA$ has only sorted access to~$\hist_{{\HighFreqSet, \LowFreqSet}}$ and does not return an item which it has not seen, if it returns an item in~$\LowFreqSet$, it needs to invoke at least~$|S| - |S| / k \in \Omega(m)$ sorted accesses.
    It follows that the expected access cost of~$\cA$ is at least~$\frac{1 - \beta - \delta - e^{-1}}{e^\eps} \cdot \Omega(m)$.
    Inequality~(\ref{ineq: lower bound on the expected sorted access cost}) follows from the assumption that~$\eps \in O(1)$, and~$\beta + \delta \le 0.6$. 
\end{proof}

\subsection{Random and Sorted Access}
\label{subsec: random sorted and sorted access}

    In this section, we present a lower bound for algorithms that can retrieve data via both random access and sorted access.

\begin{theorem}
    \label{theorem: lower bound for sorted and random access}
    Let~$\eps, \delta, \beta$ be non-negative parameters, s.t.,~$\eps \le 1, \delta + \beta \le 0.05$. 
    Let~$\cA$ be an algorithm that has both~\emph{sorted access} and~\emph{random access} to~$\hist$, does not return items which it has not seen, and for each input, with probability at least~$1 - \beta$, returns a solution that is~$\paren{n - 2, k}$-accurate.
    Then there exists a family of histograms~$\cH$, if~$\hist$ is sampled uniformly at random from~$\cH$, it holds that
    \begin{align} \small
    \vspace{-3mm}
        \label{ineq: lower bound on the expected sorted and random access cost}
        \E{\hist}{ cost \PAREN{\cA, \hist} } 
        \in 
        \Omega \PAREN{ \sqrt{mk} }. 
    \end{align}
    \vspace{-4mm}
\end{theorem}

\vspace{-4mm}
\begin{proof}
    Let~$\cH$ be the collection of all possible~$\hist_{{\HighFreqSet, \LowFreqSet}}$ generated as follows: 
    \begin{itemize}[leftmargin=4.5mm, topsep=1pt, itemsep=1pt, partopsep=1pt, parsep=1pt]
        \item 
        First, we sample an $S$ from~$\binom{[m]}{\tau}$ uniformly at random, where~$\tau \doteq \sqrt{mk}$. 
        
        \item  
        Then we sample an~$\LowFreqSet$ from~$\binom{S}{|S| / k}$ uniformly at random,
        and construct a histogram~$\hist_{{\HighFreqSet, \LowFreqSet}}$ as described by Equation~(\ref{equa: construction of random histogram based on S}).%
    \end{itemize}
    Since Lemma~\ref{lemma: probability of reporting back candiates} holds for each~$S \subseteq [m]$, 
    we have 
    \begin{align*} \small
    \begin{array}{c}
        \P{S, \LowFreqSet, \cA}{ \cA \PAREN{\hist_{\HighFreqSet, \LowFreqSet}} \cap \LowFreqSet \neq \emptyset } 
            \ge \frac{1}{e^{\eps}} \PAREN{1 - \beta - \delta - \frac{1}{e}} \\
            \stackrel{(a)}{\ge} e^{-1} \PAREN{1 - 0.05 - e^{-1}} 
            \ge 0.21, \qquad \quad 
    \end{array}
    \vspace{-1mm}
    \end{align*}
    where the randomness is first over the choice of~$S$, then over the choice of~$\LowFreqSet$, and finally over the output of~$\cA$, and inequality~$(a)$  follows from the assumption that~$\eps \le 1$ and~$\beta + \delta \le 0.05$. 
    In what follows, we omit the subscripts~$S, \LowFreqSet, \cA$ from the probability notations, when the source of randomness is clear from the context. 
    
    \vspace{-1mm}
    Consider the event~$\cE:$~$\cA$ accesses some item~$i \in \LowFreqSet$.
    As~$\cA$ does not return an item which it has not seen, the event~$\cE$ is a necessary condition for~$\cA \PAREN{\hist_{\HighFreqSet, \LowFreqSet}} \cap \LowFreqSet \neq \emptyset$. 
    Hence, 
    \vspace{-2mm}
    \begin{align*} \small
        \begin{array}{c}
            \P{\cE} 
                \ge \P{ \cA \PAREN{\hist_{\HighFreqSet, \LowFreqSet}} \cap \LowFreqSet \neq \emptyset }. 
        \end{array}
    \end{align*}
    \vspace{-5mm}
    
    Let~$\eta \doteq \tau / 20$. 
    We decompose~$\cE$ into two mutually exclusive events:~$\cE_1:$~$\cA$ accesses some item~$i \in \LowFreqSet$ for the first time within~$\eta$ access operations;
    $\cE_2: $~$\cA$ accesses some item~$i \in \LowFreqSet$ for the first time after~$\eta$ access operations.
    Then~$\P{\cE} = \P{\cE_1} + \P{\cE_2}$. 
    
    \begin{lemma}
        \label{lemma: lower bound of E1}
        The probability that~$\cA$ accesses some item~$i \in \LowFreqSet$ for the first time within~$\eta$ access operations, denoted by~$\P{\cE_1}$, is upper bounded by 
        $\P{\cE_1} \le 0.19$. 
    \end{lemma}
    
    The proof of Lemma~\ref{lemma: lower bound of E1} is omitted here, and is included in Appendix~\ref{appendix: proof for subsec: random sorted and sorted access}.
    Intuitively, the lemma holds since: 1)~$\cA$ can not access some~$i \in \LowFreqSet$ within~$\eta$ sorted accesses; 2) because of the way it is generated, ~$\LowFreqSet$ is a random subset from~$[m]$ of size~$\sqrt{mk} / k$, hence it is also unlikely for~$\cA$ to come across some~$i \in \LowFreqSet$ with at most~$\eta = \sqrt{mk} / 20$ random access. 
    To conclude the justification of~\eqref{ineq: lower bound on the expected sorted and random access cost}, and hence prove the Theorem, we apply Lemma~\ref{lemma: lower bound of E1}.
    \begin{align*} \small
        \begin{array}{ll}
            \P{\cE_2} 
                &= \P{\cE} - \P{\cE_1} \\
                &\ge \P{ \cA \PAREN{\hist_{\HighFreqSet, \LowFreqSet}} \cap \LowFreqSet \neq \emptyset } - \P{\cE_1} \\ 
                &\ge 0.21 - 0.19 
                \in \Omega (1). 
        \end{array}
    \end{align*}
    
    But when~$\cE_2$ happens, the access cost is~$\Omega(\eta)$. 
    Therefore, the expected access cost of~$\cA$ is lower bounded by~$\Omega(\eta) = \Omega(\sqrt{mk})$.%
    \end{proof}

%% file: related-work.tex
\section{Related Work}
\label{sec: related-work}

{\it Private Selection.} 
The private top-$1$ selection problem is  a special case of the private top-$k$ problem. 
The latter has been studied extensively, e.g., the exponential mechanism~\citep{McSherryT07}, report noisy max~\citep{DR14}, permute-and-flip~\citep{McKennaS20, Ding20}.
Of interest is the permute-and-flip mechanism: when the largest score of the items is known a prior , the mechanism can potentially terminate without iterating over all~$m$ items.
However, in this scenario, an asymptotic upper bound for the number of items evaluated remains unresolved.

{\it Private Top-$k$ Mechanisms.} \citeauthor{BLST10}~\citeyearpar{BLST10} were the first to apply the ``peeling exponential mechanism'', which iteratively invoked the exponential mechanism to select the item with highest score, then remove it.
They also proposed an oneshot Laplace mechanism for private top-$k$ selection.
\citeauthor{BLST10}~\citeyearpar{BLST10} analyzed the pure differential privacy guarantees of both algorithms. 
Subsequently, \citeauthor{DurfeeR19}~\citeyearpar{DurfeeR19} showed that the peeling exponential mechanism has an equivalent oneshot implementation (i.e., Algorithm~\ref{algo: top k algorithm} with Gumbel noise), and studied its approximate privacy guarantee.
~\citeauthor{QiaoSZ21}~\citeyearpar{QiaoSZ21} provided the approximate privacy guarantee for the oneshot Laplace mechanism, without the help of the composition theorem. 

Both~\citeauthor{BLST10}~\citeyearpar{BLST10} and~\citeauthor{DurfeeR19}~\citeyearpar{DurfeeR19} have proposed private algorithms which estimate top-$k$ based on the true top~$\bar{k}$ items for some~$\bar{k} \ge k$.
Given an integer~$k$, both algorithm may need to set~$\bar{k} = m$, in order to return~$k$ items. 

{\it Accuracy Lower Bound.}
\citeauthor{BafnaU17}~\citeyearpar{BafnaU17} and~\citeauthor{SteinkeU17}~\citeyearpar{SteinkeU17} show that, for  approximate private algorithms, the error guarantees of existing algorithms~\citep{McSherryT07, DR14, DurfeeR19, QiaoSZ21} are essentially optimal.

%% file: conclusion.tex
\section{Conclusions and Future Directions}
\label{sec: summary}

    In this paper, we systematically advance our understanding of the access cost of private top-$k$ selection algorithm. 
    We introduce the first algorithm with sublinear access cost, and provide lower bounds for three access models, showing that supporting both sorted access and random access is the key to breaking the linear access cost barrier, and that the access cost of our algorithm is optimal. 

    We believe our work is a first step towards a comprehensive study of building a differentially private top-$k$ algorithm on top of existing data analytics systems. 
    Our focus in this work is primarily on advancing theoretical understanding of the problem, %
    assuming that sorted access and random access operations have the same cost.
    Interesting future directions include conducting empirical evaluations, and investigating scenarios where the costs of these two operations differ.

%% file: acknowledgement.tex
\section*{Acknowledgements}

We thank the anonymous reviewers for their constructive feedback, which has helped us to improve our manuscript.
In particular, we acknowledge the suggestion that our algorithm can be applied to the exponential mechanism, resulting in an~$O(\sqrt{m})$ expected access cost. 

 Hao Wu is supported by an Australian Government Research Training Program (RTP) Scholarship.

%% file: appendix.tex
\section{Probability Inequalities}

\begin{fact}[Chernoff Bound, (\citeauthor{MU17}~\citeyear{MU17}, Theorem 4.4 and 4.5)]
    \label{fact: chernoff bound sampling with replacement}
    Let $X_1, X_2, \ldots, X_n$ be independent Poisson trials  %
    such that, for~$i \in [m]$, $\P{X_i = 1} = p_i$, where~$0 < p_i < 1$. Then, for~$X = \sum_{i = 1}^n X_i$,  $\mu = \E{X}$,
    \begin{align}
        \P{ X \ge (1 + \lambda) \mu } 
            &\le \PAREN{ \frac{ e^{\lambda} }{ \paren{1 + \lambda}^{1 + \lambda} } }^{\mu},  
            &\forall \lambda > 0, \\
        \P{ X \ge (1 + \lambda) \mu } 
            &\le \PAREN{ \frac{ e^{\lambda} }{ \paren{1 + \lambda}^{1 + \lambda} } }^{\mu}
            \le e^{ - \lambda^2 \mu / 3},
            &\forall \lambda \in (0, 1], \\
        \P{ X \le (1 - \lambda) \mu } 
            &\le \PAREN{ \frac{ e^{-\lambda} }{ \paren{1 - \lambda}^{1 - \lambda} } }^{\mu}
            \le e^{ - \lambda^2 \mu / 2},
            &\forall \lambda \in \paren{0, 1}. 
    \end{align}
\end{fact}

The known concentration inequalities for sampling with replacement can be transferred to the case of sampling without replacement, based on a notable reduction technique. 

\begin{fact}[\citep{Hoeffding1994}  ]
    \label{fact: Hoeffing reduction of momenting generating function}
    Let $\cX = (x_1, \ldots, x_N)$ be a finite population of~$N$ real points,~$Y_1, \ldots, Y_n$ denote a random sample without replacement from~$\cX$ and~$X_1, \ldots, X_n$ denote a random
    sample with replacement from~$\cX$. 
    If~$f : \R \rightarrow \R$ is continuous and convex, then
    $$
        \E{ f \PAREN{\sum_{i = 1}^n Y_i} }
        \le 
        \E{ f \PAREN{\sum_{i = 1}^n X_i} }.
    $$
\end{fact}

In particular, the lower bound presented in Fact~\ref{fact: chernoff bound sampling with replacement} can be converted into the following one, by combining 
its proof in~\citep{MU17} and Fact~\ref{fact: Hoeffing reduction of momenting generating function}.

\begin{corollary}[Chernoff bound]
    \label{fact: chernoff bound sampling without replacement}
    Let $\cX = (x_1, \ldots, x_N) \in \set{0, 1}^N$ be a finite population of $N$ binary points and $Y_1, \ldots, Y_n$ be a random sample drawn without replacement from the population. 
    Then, for~$Y = \sum_{i = 1}^n Y_i$,
    \begin{align}
        \label{ineq: chernoff lower bound sampling without replacement}
        \P{ \sum_{i = 1}^n Y_i \le (1 - \lambda) \mu } 
            \le \PAREN{ \frac{ e^{-\lambda} }{ \paren{1 - \lambda}^{1 - \lambda} } }^{\mu}
            \le e^{ - \lambda^2 \mu / 2},
            \qquad \forall \lambda \in [0, 1). 
    \end{align}
    where $\mu = n p$ is the expectation of~$\sum_{i = 1}^n Y_i$, and $p \doteq \frac{1}{N} \sum_{i = 1}^N x_i$ is the mean of $\cX$.
\end{corollary}

Note that, in Corollary~\ref{fact: chernoff bound sampling without replacement}, we also extend the range of~$\lambda$ from~$\set{0, 1}$, to~$[0, 1)$.
When~$\lambda = 0$,~
$
    \PAREN{ \frac{ e^{-\lambda} }{ \paren{1 - \lambda}^{1 - \lambda} } }^{\mu} 
        = e^{ - \lambda^2 \mu / 2}
        = 1, 
$
and Inequality~(\ref{ineq: chernoff lower bound sampling without replacement}) holds trivially.

\begin{fact}[\citep{BlitzsteinH14}]
    \label{fact: equations of gamma function}
    Let~$\GammaFunt{a} \doteq \int_{0}^{\infty } x^{a - 1} e^{-x} \, dx$, $\forall a > 0$ be the \emph{Gamma function}.
    It holds that~$\GammaFunt{a + 1} = a \GammaFunt{a}$, and~$\GammaFunt{\tfrac{1}{2}} = \sqrt{\pi}$.
    Then for each~$k \in \N^+$, we have~$\GammaFunt{k} = \paren{k - 1}!$ and
    \begin{align}
        \Gamma \PAREN{\frac{1}{2} + k }
            &= \PAREN{k - \frac{1}{2}} \PAREN{k - \frac{3}{2}} \cdots \frac{1}{2} \Gamma \PAREN{\frac{1}{2}}
            = \frac{2k - 1}{2} \frac{2k - 3}{2} \cdots \frac{1}{2} \sqrt{\pi} 
            = \frac{(2k)!}{4^{k} k!} \sqrt{\pi}.
    \end{align}
\end{fact}

\begin{fact}[\citep{Ross2018}]
    \label{fact: gamma and beta}
    Given \emph{shape parameters}~$\alpha, \beta > 0$, define \emph{beta function}~$\BetaFunt{\alpha}{\beta} \doteq \int_0^1 x^{\alpha - 1} \paren{1 - x}^{\beta - 1} \, dx$.
    Then $\BetaFunt{\alpha}{\beta} =  \frac{\Gamma(\alpha) \Gamma(\beta)}{\Gamma(\alpha + \beta)}$, 
    In particular,
    \begin{align}
        \BetaFunt{\alpha}{\beta} = \frac{ \paren{\alpha - 1}! \paren{\beta - 1}! }{\paren{\alpha + \beta - 1}!}, 
        &\qquad \forall \alpha, \beta \in \N^+.     
    \end{align}
\end{fact}

The factorials can be estimated as follows.

\begin{fact}[Stirling's Approximation \cite{R55, MN09}] \label{fact: stirling} For $k = 1, 2, ...$
    \begin{equation} \label{ineq: stirling}
            \sqrt{2 \pi k} \left( \frac{k}{e} \right)^k \exp \left( \frac{1}{12k + 1} \right) 
            \le k! 
            \le 
            \sqrt{2 \pi k} \left( \frac{k}{e} \right)^k \exp \left(  \frac{1}{12k} \right).
    \end{equation}
\end{fact}

\section{Proofs for Section~\ref{sec: solution}}

\subsection{Proofs for Section~\ref{subsec: starwman solution}}
\label{appendix: proof for subsec: starwman solution}

\begin{proof}[Proof for Lemma~\ref{lemma: permutaion and independence}]
    For each~$i \in [m]$, let~$p_{{\textstyle\mathstrut} Z_{i} } \paren{ \cdot }$ be the density function of random variable~$Z_i$. 
    Since the~$Z_i$'s are i.i.d. random variables, they share the same density function, i.e.,~$p_{{\textstyle\mathstrut} Z_{1} } \paren{ \cdot } = \cdots = p_{{\textstyle\mathstrut} Z_{m} } \paren{ \cdot }$.
    
    Let~$S_m$ be the collection of all possible permutations over~$[m]$, and~$s$ refer to a permutation in~$S_m$  . 
    Further, let~$z_{(1)} \ge \ldots \ge z_{(m)}$  denote a possible realization of~$Z_{(1)}, \ldots, Z_{(m)}$, and~$z_{(1: m)} \doteq \paren{ z_{(1)}, \ldots, z_{(m)} }$.
    We write~$I_{(1:m)} = s$, if~$I_{(j)} = s(j), \forall j \in [m]$, and~$Z_{(1:m)} = z_{(1:m)}$, when~$Z_{(j)} = z_{(j)}, \forall j \in [m]$.

    Let~$p_{ {\textstyle\mathstrut} I_{(1 : m)}, Z_{(1: m)} } \paren{ s, z_{(1:m)} }$ be the probability density  when~$I_{(1:m)} = s$ and~$Z_{(1:m)} = z_{(1:m)}$.
    The probability density,~$p_{ {\textstyle\mathstrut} Z_{(1: m)} } \paren{ z_{(1: m)} }$ of~$Z_{(1:m)} = z_{(1:m)}$ is given by
    \begin{align*}
        p_{ {\textstyle\mathstrut} Z_{(1 : m)} } \paren{ z_{(1: m)} }
            &= \sum_{ s \in S_m } p_{ {\textstyle\mathstrut} I_{(1 : m)}, Z_{(1: m)} } \paren{ s, z_{(1:m)} } \\
            &= \sum_{ s \in S_m } \prod_{j \in [m]} p_{{\textstyle\mathstrut} Z_{s(j)}} \paren{ z_{(j)} } \\
            &= \sum_{ s \in S_m } \prod_{j \in [m]} p_{{\textstyle\mathstrut} Z_{1}} \paren{ z_{(j)} } \\
            &= m! \prod_{j \in [m]} p_{{\textstyle\mathstrut} Z_{1}} \paren{ z_{(j)} }. 
    \end{align*}
    Hence, for a given~$s \in S_m$, the probability density of~$I_{(1:m)} = s$, conditioned on~$Z_{(1:m)} = z_{(1:m)}$, is given by
    \begin{align*}
        p_{ I_{(1:m)} \mid {\textstyle\mathstrut} Z_{(1 : m)} } \paren{ s \mid z_{(1:m)} }
            = \frac{
                p_{ {\textstyle\mathstrut} I_{(1 : m)}, Z_{(1: m)} } \paren{ s, z_{(1:m)} }
            }{
                p_{ {\textstyle\mathstrut} Z_{(1 : m)} } \paren{ z_{(1: m)} }
            } 
            = \frac{
                \prod_{j \in [m]} p_{{\textstyle\mathstrut} Z_{1}} \paren{ z_{(j)} }
            }{
                m! \prod_{j \in [m]} p_{{\textstyle\mathstrut} Z_{1}} \paren{ z_{(j)} }
            } = \frac{1}{m!}\,, 
    \end{align*}
    which is independent of the values of the~$Z_{(1:m)}$. 
    Finally, 
    \begin{align*}
        p_{ {\textstyle\mathstrut} I_{(1:m)} } \paren{ s }
        =
        \E{Z_{(1 : m)}} { p_{ I_{(1:m)} \mid {\textstyle\mathstrut} Z_{(1 : m)} } \paren{ s \mid Z_{(1:m)} } }
        = \frac{1}{m!}.
    \end{align*}
\end{proof}

\subsubsection{Proof of Theorem~\ref{theorem: access cost of algo: modified threshold algorithm}}

Before the proof of Theorem~\ref{theorem: access cost of algo: modified threshold algorithm},
we present two supporting lemmas. 

\begin{lemma}
    \label{lemma: probability upper bound 1 for access cost of algo: modified threshold algorithm}
    Let~$\AccessCost{\algoPrivTA}{\DataList{1}}$ be the access cost of Algorithm~\ref{algo: modified threshold algorithm}.
    Then 
    \begin{align}
        \P{
            \AccessCost{\algoPrivTA}{\DataList{1}} \ge 2 \cdot r
        }
        \le \frac{e^k}{k^k} \cdot \frac{ \paren{r^2 / m}^k }{ e^{r^2 / m} }, 
        & \qquad  \forall r \in \N,\, s.t.,\, r \ge \sqrt{mk}.
    \end{align}
\end{lemma}

\begin{proof}[Proof for Lemma~\ref{lemma: probability upper bound 1 for access cost of algo: modified threshold algorithm}]
    Let~$\cS_r \doteq \set{ i_{(1)}, \ldots, i_{(r)} }$ be the top-$r$ items with highest scores in~$\hist$. 
    Also consider the items in the first~$r$ tuples in the array~$\DataList{2}$, denoted by $I_{(1:r)} = \set{ I_{(1)}, \ldots, I_{(r)} }$.
    
    Consider the following event: 
    $$
        \cE \doteq \card{ \cS_r  \cap I_{(1:r)} } > k\,.
    $$
    We claim that
    \begin{enumerate}
        \item When event~$\cE$ happens,~$\AccessCost{\algoPrivTA}{\DataList{1}} < 2r$.
        \item The complement of~$\cE$, denoted by~$\bar{\cE}$, happens with probability~$\P{\Bar{\cE}} \le \frac{e^k}{k^k} \cdot \frac{ \paren{r^2 / m}^k }{ e^{r^2 / m} }$.
    \end{enumerate}
    Combing both claims, we get
    \begin{align*}
        \P{\AccessCost{\algoPrivTA}{\DataList{1}} \ge 2r} 
            \le \P{\Bar{\cE}} 
            \le \frac{e^k}{k^k} \cdot \frac{ \paren{r^2 / m}^k }{ e^{r^2 / m} }.
    \end{align*}
    
    {\it Claim One.}
    It suffices to show that, when event~$\cE$ happens, Algorithm~\ref{algo: threshold algorithm} stops before~$r$ rounds 
    , where each round involves executing lines~\ref{line: threshold algorithm round start} to~\ref{line: threshold algorithm round end} in Algorithm~\ref{algo: threshold algorithm}.
    In such a case, the number of random accesses incurred (to~$\hist$) is less than~$r$. 
    Therefore, the total number of access cost to~$\hist$ is less than~$2 r$. 
    
    Assume that Algorithm~\ref{algo: threshold algorithm} runs for~$r$ rounds.
    Since~$i_{(r)}$ and~$I_{(r)}$ are the last encountered items, the corresponding threshold (Algorithm~\ref{algo: threshold algorithm}, Line~\ref{line: threshold}) is given by~$\tau_r \doteq \AggregationFunt(\hist[i_{(r)}], Z_{I_{(r)}})$. 
    
    But for each~$i \in \set{ i_{(1)}, \ldots, i_{(r)}  } \cap I_{(1:r)}$, it holds that~$\hist[i] \ge \hist[i_{(r)}]$, and~$Z_i \ge Z_{I_{(r)}}$.
    Since~$\AggregationFunt$ is monotone, the score of item~$i$,~$\AggregationFunt(\hist[i], Z_{i})$, is at least~$\tau_r$.
    Since event~$\cE$ happens, at least~$k$ items encountered by the algorithm have score at least~$\tau$.
    Therefore, the stopping condition of Algorithm~\ref{algo: threshold algorithm} should be satisfied.
    
    {\it Claim Two.}
    Via Lemma~\ref{lemma: permutaion and independence}, the~$I_{(1)}, \ldots, I_{(m)}$ distribute uniformly over all permutations of $[m]$.
    Therefore, the set~$I_{(1:r)}$ can be viewed as a uniform sample (without replacement) of~$r$ elements from~$[m]$.
    Hence, for each $j \in [r]$, the probability that~$i_{(j)}$ belongs to $I_{(1:r)}$ is given by
    $$
        \P{i_{(j)} \in I_{(1:r)} } = r / m\,. 
    $$ 
    Let~$\indicator{i_{(j)} \in I_{(1:r)}}$ be the indicator for the event that~$i_{(j)} \in I_{(1:r)}$, and~$Y \doteq \sum_{j \in [r]} \indicator{i_{(j)} \in I_{(1:r)} }$. 
    The event~$\bar{\cE}$ is equivalent to~$Y \le k$. 
    Observe that
    \begin{align}
        &\P{ \indicator{i_{(j)} \in I_{(1:r)} } = 1 } 
        = \P{i_{(j)} \in I_{(1:r)} } = r / m, \\
        &\mu \doteq \E{ Y } = \sum_{j \in [r] } \E{ \indicator{i_{(j)} \in I_{(1:r)} } } = r^2 / m \ge k. 
    \end{align}
    Since~$k / \mu \in (0, 1]$,~$\lambda \doteq  1 - k / \mu \in [0, 1)$.
    By the Chernoff bound for sampling without replacement (Fact~\ref{fact: chernoff bound sampling without replacement}), 
    we have 
    \begin{align*}
        \P{ Y \le k } 
            = \P{ Y \le \PAREN{1 - \lambda} \cdot \mu } 
            \le \PAREN{ \frac{ e^{-\lambda} }{ \paren{1 - \lambda}^{1 - \lambda} } }^{\mu} 
            = \PAREN{ \frac{ e^{k / \mu - 1} }{ \paren{k / \mu}^{k / \mu} } }^{\mu}
            = \frac{ e^{k - \mu} }{ \paren{k / \mu}^{k} }
            = \frac{ e^k \mu^k }{ k^k e^\mu }
            = \frac{e^k}{k^k} \cdot \frac{ \paren{r^2 / m}^k }{ e^{r^2 / m} }.
    \end{align*} 
\end{proof}

\begin{lemma}
    \label{lemma: probability upper bound 2 for access cost of algo: modified threshold algorithm}
    Let~$\AccessCost{\algoPrivTA}{\DataList{1}}$ be the access cost of Algorithm~\ref{algo: modified threshold algorithm}.
    Then 
    \begin{align}
        \P{
            \AccessCost{\algoPrivTA}{\DataList{1}} \ge 2 \cdot r
        }
        \le \frac{e^k}{k^k} \cdot \frac{ \paren{r^2 / m}^k }{ e^{r^2 / m} }, 
        & \qquad  \forall r \in \R,\, s.t.,\, r \ge \sqrt{mk}.
    \end{align}
\end{lemma}

\begin{proof}[Proof for Lemma~\ref{lemma: probability upper bound 2 for access cost of algo: modified threshold algorithm}]
    Since Algorithm~\ref{algo: modified threshold algorithm} executes  the threshold algorithm (Algorithm~\ref{algo: threshold algorithm}) with two sorted arrays~$\DataList{1}$ and~$\DataList{2}$, the access cost of Algorithm~\ref{algo: modified threshold algorithm} on~$\DataList{1}$, denoted by~$\AccessCost{\algoPrivTA}{\DataList{1}}$, is an even integer due to the way the threshold algorithm operates.
    Therefore, the event~$\AccessCost{\algoPrivTA}{\DataList{1}} \ge 2 \cdot r$ is equivalent to the event~$\AccessCost{\algoPrivTA}{\DataList{1}} \ge \lceil 2 \cdot r \rceil_{even}$, where~$\lceil 2 \cdot r \rceil_{even}$ denotes the smallest even integer that is at least~$2 \cdot r$.

    Define~$r_\star \doteq \tfrac{1}{2} \cdot \lceil 2 \cdot r \rceil_{even} \in \N$.
    Clearly it holds that~$r_\star \ge r$.
    Via Lemma~\ref{lemma: probability upper bound 1 for access cost of algo: modified threshold algorithm}, for each~$r \in \R,\, s.t.,\, r \ge \sqrt{mk}$, 
    \begin{align*}
        \P{
            \AccessCost{\algoPrivTA}{\DataList{1}} \ge 2 \cdot r
        }
        &= \P{
            \AccessCost{\algoPrivTA}{\DataList{1}} \ge \lceil 2 \cdot r \rceil_{even}
        } \\
        &= \P{
            \AccessCost{\algoPrivTA}{\DataList{1}} \ge 2 \cdot r_\star
        } 
        \le \frac{e^k}{k^k} \cdot \frac{ \paren{r_\star^2 / m}^k }{ e^{ r_\star^2 / m} }.
    \end{align*}

    Consider the function~$y \doteq x^k / e^x$. 
    As~$y' = \tfrac{k x^{k - 1} e^x - x^k e^x}{e^{2x}}$, $y$ is decreasing when~$x \ge k$.
    Noting that~$r_\star^2 / m \ge r^2 / m$, we have
    \begin{align*}
        \P{
            \AccessCost{\algoPrivTA}{\DataList{1}} \ge 2 \cdot r
        }
        \le \frac{e^k}{k^k} \cdot \frac{ \paren{r_\star^2 / m}^k }{ e^{ r_\star^2 / m} }
        \le \frac{e^k}{k^k} \cdot \frac{ \paren{r^2 / m}^k }{ e^{ r^2 / m} },
        \qquad  \forall r \in \R,\, s.t.,\, r \ge \sqrt{mk}.
    \end{align*}
\end{proof}

We are now ready to prove Theorem~\ref{theorem: access cost of algo: modified threshold algorithm}.

\begin{proof}[Proof of Theorem~\ref{theorem: access cost of algo: modified threshold algorithm}]

    First, we can rewrite    
    \begin{align*}
            \E{
                \AccessCost{\algoPrivTA}{\DataList{1}}
            }
            = \int_0^\infty \P{ \AccessCost{\algoPrivTA}{\DataList{1}} \ge s } \, ds 
            = 2 \cdot \int_0^\infty \P{ \AccessCost{\algoPrivTA}{\DataList{1}} \ge 2 \cdot r } \, dr,
    \end{align*}
    where the last inequality follows from a change of variable of~$r \doteq s / 2$.
    
    Decomposing the integral further, we have 
    \begin{align*}
        \int_0^\infty \P{ \AccessCost{\algoPrivTA}{\DataList{1}} \ge 2  r } \, dr 
            &= \int_0^{ \sqrt{mk} } \P{ \AccessCost{\algoPrivTA}{\DataList{1}} \ge 2 r } \, dr 
            + \int_{ \sqrt{mk} }^\infty \P{ \AccessCost{\algoPrivTA}{\DataList{1}} \ge 2 r } \, dr \\
            &\le \sqrt{mk} 
            + \int_{ \sqrt{mk} }^\infty \P{ \AccessCost{\algoPrivTA}{\DataList{1}} \ge 2 r } \, dr.
    \end{align*}
    
    Via Lemma~\ref{lemma: probability upper bound 2 for access cost of algo: modified threshold algorithm}, we can bound the last integral by    
    \begin{align*}
        \int_{ \sqrt{mk} }^\infty \P{ \AccessCost{\algoPrivTA}{\DataList{1}} \ge 2 r } \, dr 
            \le \int_{ \sqrt{mk} }^\infty \frac{e^k}{k^k} \cdot \frac{ \paren{ r^2 / m}^k }{ e^{ r^2 / m} } \, dr 
            = \int_{ k }^\infty \frac{e^k}{k^k} \cdot t^k e^{-k} \, \sqrt{m} \frac{dt }{2 \sqrt{t}}, 
    \end{align*}
    where the last inequality follows from a change of variable of~$t \doteq r^2 / m$.
    Via the definition and property of Gamma function (Fact~\ref{fact: equations of gamma function}), 
    \begin{align*}
        \int_{ k }^\infty \frac{e^k}{k^k} \cdot t^k e^{-k} \, \sqrt{m} \frac{dt }{2 \sqrt{t}}
            &\le \frac{\sqrt{m}}{2} \cdot \frac{e^k}{k^k} \cdot \int_0^\infty t^{k - 1 / 2} e^{-k} \, dt \\
            &= \frac{\sqrt{m}}{2} \cdot \frac{e^k}{k^k} \cdot \GammaFunt{k + 1 / 2} 
            = \frac{\sqrt{m}}{2} \cdot \frac{e^k}{k^k} \cdot \frac{\paren{2k}!}{4^k k!}.
    \end{align*}

    Finally, by Stirling's approximation (Fact~\ref{fact: stirling}),
    \begin{align*}
        \frac{e^k}{k^k} \cdot
        \frac{\paren{2k}!}{4^k k!}
            &\le \frac{e^k}{k^k} \cdot
            \frac{ 
                \sqrt{2 \pi \cdot 2k} \PAREN{\frac{2k}{e}}^{2k} 
                \exp \PAREN{ \frac{1}{12 \cdot 2k} }
            }{
                4^k \sqrt{2 \pi \cdot k} \PAREN{\frac{k}{e}}^{k}
                \exp \PAREN{ \frac{1}{12 \cdot k + 1} }
            }
            \le \sqrt{2}.
    \end{align*}

    Combing the previous inequalities, we show that 
    \begin{align*}
        \E{
            \AccessCost{\algoPrivTA}{\DataList{1}}
        }
        \le 2 \cdot \PAREN{ \sqrt{mk} + \frac{\sqrt{m}}{\sqrt{2}} },
    \end{align*}
    which proves the theorem.
\end{proof}

\subsection{Proofs for Section~\ref{subsec: An Online Sampling Approach}}
\label{appendix: proof for subsec: An Online Sampling Approach}

\begin{proof}[Proof of Theorem~\ref{theorem: oracle}]
    This is a directly consequence of the facts that 
    \begin{itemize}
        \item Based on a technique by~\citeauthor{BrassardK88}~\citeyearpar{BrassardK88} for sampling a random perturbation on the fly, and on the discussions in Section~\ref{subsec: An Online Sampling Approach}, it holds that each~$I_{(j)}, j \in [m]$, and each value of inverted index~$\invertedIndex{2}(i), i \in [m]$ can be sampled on the fly in~$O(1)$ times.
        
        \item Based on Lemma~\ref{lemma: sampling zj}, each~$Z_{(j)}, j \in [m]$ can be sampled on the fly with~$O(\log \log m)$ expected time. 
    \end{itemize}
\end{proof}

\begin{proof}[Proof of Fact~\ref{fact: inversion method for order statistics}.]
    Recall that~$F$ is the cumulative distribution function of~$Z_1, \ldots, Z_m$, and~$U_1, \ldots, U_m$ are independent uniform random variables on~$[0, 1]$.
    We compare the two post-processing procedures: 
    \begin{itemize}[leftmargin=4.5mm, topsep=1pt, itemsep=1pt, partopsep=1pt, parsep=1pt]
        \item By Fact~\ref{fact: random varaible generation by uniform}, we can obtain~$Z_1, \ldots, Z_m$ by computing~$F^{-1}(U_1), \ldots, F^{-1}(U_m)$; then we obtain~$Z_{(1)}, \ldots, Z_{(m)}$ by sorting this sequence in descending order.
    
        \item Alternatively, we first sort~$U_1, \ldots, U_m$ in descending order, to obtain a sequence~$\uniOrderStat{1}$, $\ldots,$ $\uniOrderStat{m}$; then we compute~$F^{-1}(U_{(1)}), \ldots, F^{-1}(U_{(m)})$.
    \end{itemize}

    \begin{figure}[!h]
    	\includegraphics[width=0.5\linewidth,center]{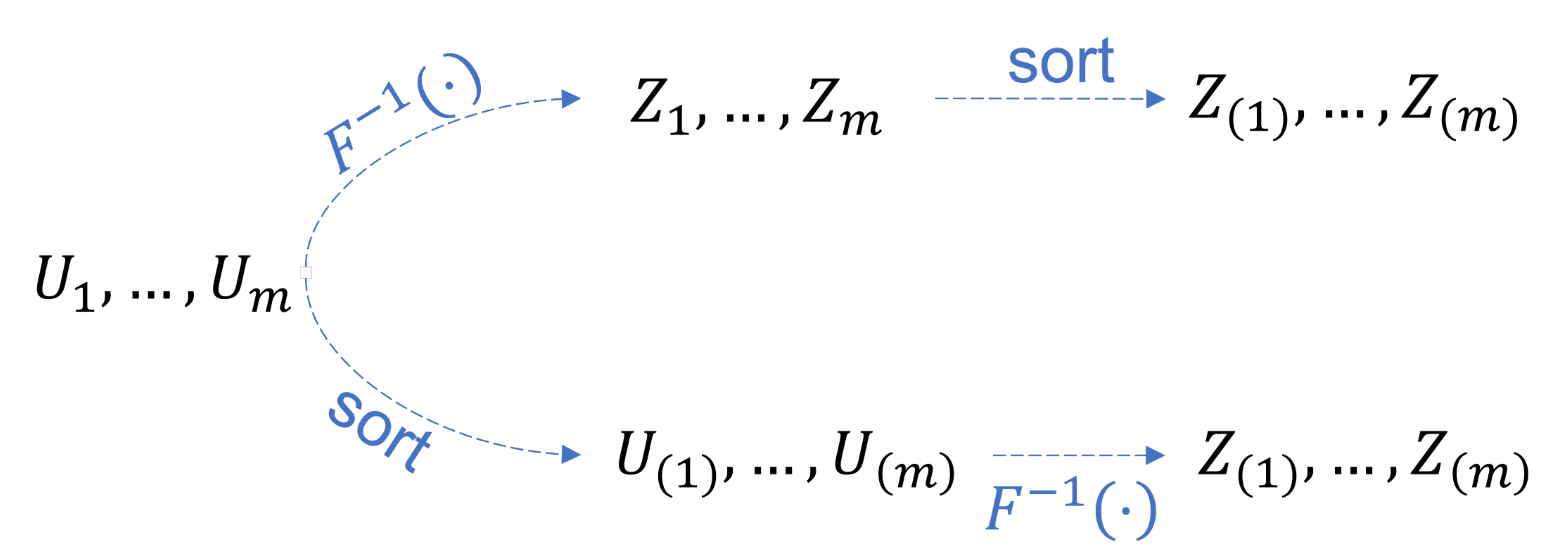}
    	\vspace{-4mm}
    	\caption{A pictorial comparison of the two procedures.}
    	\label{fig:order statistics}
    \end{figure}

    We claim that the two procedures output the same sequence.

    First, observe that the following two multisets are the same~$\set{ F^{-1}(U_1), \ldots, F^{-1}(U_m) }$ and~$\set{ F^{-1}(U_{(1)}), \ldots, F^{-1}(U_{(m)}) }$.
    By construction, $Z_{(1)}, \ldots, Z_{(m)}$ is a sorted sequence of the multiset.
    To prove the claim, it suffices to show that~$F^{-1}(U_{(1)}), \ldots, F^{-1}(U_{(m)})$ is also a sorted sequence.
    This is true: since~$F$ is a cumulative distribution function on~$\R$, it is non-decreasing; it follows that for every two~$1 \le j <  j' \le m$, $U_{(j)} \ge U_{(j')}$ implies that $F^{-1}(U_{(j)}) \ge F^{-1}(U_{(j')})$.
\end{proof}

\begin{proof}[Proof of Theorem~\ref{theorem: converting beta distribitons}]

In this proof, we write the density function of~$X$ as~$p_X (\cdot)$ and the density function of~$Y$ as~$p_Y (\cdot)$, to distinguish between the two.

If~$X \sim \BetaDist{m - j + 1}{j}$, then its density function satisfies
$$
    p_X (x) = \frac{\paren{x}^{m - j} \paren{1 - x}^{j - 1}}{\BetaFunt{m - j + 1}{j}}, 
    \qquad
    \forall x \in [0, 1].
$$
Based on Fact~\ref{fact: gamma and beta},~$\BetaFunt{m - j + 1}{j} = \frac{\paren{m - j}! \paren{j - 1}!}{m!}$, which proves the first claim.

Secondly, if~$X \sim \BetaDist{m - j + 1}{j - \ell}$, then the density of~$Y = u_{(\ell)} \cdot X$, 
is given by, 
\begin{align*}
    p_Y (y) = p_X \PAREN{ \frac{y}{u_{(\ell)}} } \cdot \frac{d X}{d Y} 
        = \frac{1}{ \BetaFunt{m - j + 1}{j - \ell} } \cdot 
        {
            \PAREN{ \frac{y}{u_{(\ell)}} }^{m - j} 
            \PAREN{1 - \PAREN{ \frac{y}{u_{(\ell)}} } }^{j - \ell - 1}
        } \cdot \frac{1}{u_{(\ell)}}\,, 
\end{align*}
Noting that~$\BetaFunt{m - j + 1}{j - \ell} = \frac{\paren{m - j}! \paren{j - \ell - 1}!}{\paren{m - \ell}!}$ proves the second claim.

The third claim follows from similar argument as the second one. 
    
\end{proof}

\subsection{Proof for Theorem~\ref{theorem: conditional distribution of the order statistics}}
\label{appendix: proof for theorem: conditional distribution of the order statistics}

We need the following fact. 

\begin{fact}[\citep{Guntuboyina2019}]
    The joint density of~$\uniOrderStat{1}, \ldots, \uniOrderStat{m}$ is given by
    \begin{align}
        p(u_{(1)}, u_{(2)}, \ldots, u_{(m)}) = m!\,, 
    \end{align}
    where~$1 \ge u_{(1)} \ge u_{(2)} \cdots \ge u_{(m)} \ge 0$.
\end{fact}

As a sanity check, note that the simplex~$\Delta_n^* \doteq \set{ u_{(1)}, u_{(2)} \ldots, u_{(m)} \in \R^m : 1 \ge u_{(1)} \ge u_{(2)} \ldots \ge u_{(m)} \ge 0}$ has volume~$1 / m!$.
Since~$\uniOrderStat{1}, \ldots, \uniOrderStat{m}$ distribute uniformly over~$\Delta_n^*$, each point of the simplex has density~$m!$. 
Based on this fact, we can derive the following lemma.

\begin{lemma}
    \label{lemma: joint density of order statistics}
    Let~$j_1, \ldots, j_t \in [m]$ be an increasing sequence of indexes.
    Then the joint distribution of~$U_{(j_1)}, \ldots, U_{(j_t)}$ is given by
    \begin{align}
        \label{equa: joint density of order statistics}
        p \PAREN{ u_{(j_1)}, \ldots, u_{(j_t)} }
            = 
            m! \cdot \frac{
                \PAREN{u_{(j_t)}}^{m - j_t}
            }{
                \PAREN{m - j_t}!
            } \cdot \frac{
                \PAREN{1 - u_{(j_1)}}^{j_1 - 1}
            }{
                \paren{j_1 - 1}!
            } \cdot 
                \prod_{s = 2}^t \frac{
                    \PAREN{u_{(j_{s - 1})} - u_{(j_s)}}^{j_s - j_{s - 1} - 1}
                }{
                    \paren{j_s - j_{s - 1} - 1}!
                }
            .
    \end{align}
\end{lemma}

\begin{proof}[Proof of Lemma]
    $p \PAREN{ u_{(j_1)}, \ldots, u_{(j_t)} }$ is given by the following integration: 
    \begin{align*}
        \int_0^1 \cdots \int_0^1
            m! \cdot \indicator{0 \le u_{(m)} \le  \ldots \le u_{(1)} \le 1}
            \, d_{u_{(m)}} \ldots d_{u_{(1 + j_t)}}
            \, d_{u_{(-1 + j_t)}} \ldots d_{u_{(1 + j_{t - 1})}} 
            \ldots
            \, d_{u_{(-1 + j_1)}} \ldots d_{u_{(1)}}. 
    \end{align*}
    Integrating with respect to~$u_{(m)}$ over the range~$[0, u_{(m - 1)}]$, we have 
    \begin{align*}
        \int_0^1 \cdots \int_0^1
            m! \cdot u_{(m - 1)} \cdot \indicator{0 \le u_{(m - 1)} \le  \ldots \le u_{(1)} \le 1}
            \, d_{u_{(m - 1)}} \ldots d_{u_{(1 + j_t)}}
            \, d_{u_{(-1 + j_t)}} \ldots d_{u_{(1 + j_{t - 1})}} 
            \ldots
            \, d_{u_{(-1 + j_1)}} \ldots d_{u_{(1)}}. 
    \end{align*}
    
    Then integrate~$u_{(m - 1)}$ over the range~$[0, u_{(m - 2)}]$, all the way down to the integral with respect to~$u_{(1 + j_t)}$ over the range~$[0, u_{(j_t)}]$.
    We obtain    
    \begin{align*}
         \int_0^1 \cdots \int_0^1
            m! \cdot \frac{\PAREN{u_{(j_t)}}^{m - j_t}}{\PAREN{m - j_t}!} \cdot 
            \indicator{0 \le u_{(j_t)} \le u_{(-1 + j_t)} \ldots \le u_{(1)} \le 1}
            \, d_{u_{(-1 + j_t)}} \ldots d_{u_{(1 + j_{t - 1})}}
            \ldots
            \, d_{u_{(-1 + j_1)}} \ldots d_{u_{(1)}}. 
    \end{align*}
    
    Integrating with respect to~$u_{(-1 + j_t)}$ over the range~$[u_{(j_t)}, u_{(- 2 + j_t)}]$, we have 
    \begin{align*}
         \int_0^1 \cdots \int_0^1
            m! \frac{\PAREN{u_{(j_t)}}^{m - j_t}}{\PAREN{m - j_t}!}
            \PAREN{u_{(-2 + j_t)} - u_{(j_t)}} 
            \cdot \indicator{0 \le u_{(j_t)} \le u_{(-2 + j_t)} \ldots \le u_{(1)} \le 1}
            \, d_{u_{(-2 + j_t)}} \ldots d_{u_{(1 + j_{t - 1})}}
            \ldots
            \, d_{u_{(-1 + j_1)}} \ldots d_{u_{(1)}}. 
    \end{align*}
    Then integrate~$u_{(- 2 + j_t)}$ over the range~$[u_{(j_t)}, u_{(-3 + j_t)}]$, all the way down to~$u_{(1 + j_{t - 1})}$ over the range~$[u_{(j_t)}, u_{(j_{t - 1})}]$:
    
    \begin{align*}
         \int_0^1 \cdots \int_0^1
            \set{
                \begin{aligned}
                     m! \cdot &\frac{\PAREN{u_{(j_t)}}^{m - j_t}}{\PAREN{m - j_t}!} \cdot 
                    \frac{\PAREN{u_{(j_{t - 1})} - u_{(j_t)}}^{ j_t - j_{t - 1} - 1 }}{\PAREN{j_t - j_{t - 1} - 1}!} \\
                    &\cdot \indicator{0 \le u_{(j_{t - 1})} \le u_{(-1 + j_{t - 1})} \ldots \le u_{(1)} \le 1} 
                \end{aligned}
            }
            \, d_{u_{(-1 + j_{t - 1})}} \ldots d_{u_{(1 + j_{t - 2})}}
            \ldots
            \, d_{u_{(-1 + j_1)}} \ldots d_{u_{(1)}}. 
    \end{align*}
    
    Repeating the above efforts proves Equation~(\ref{equa: joint density of order statistics}). 
\end{proof}

\begin{proof}[Proof of Theorem~\ref{theorem: conditional distribution of the order statistics}]
    \hfill
    
    Claim One: 
    Applying Lemma~\ref{lemma: joint density of order statistics} directly, we have 
    \begin{align*}
        p \PAREN{ u_{(j)} } 
            &= 
            m! \cdot \frac{
                \PAREN{u_{(j)}}^{m - j}
            }{
                \PAREN{m - j}!
            } \cdot \frac{
                \PAREN{1 - u_{(j)}}^{j - 1}
            }{
                \PAREN{j - 1}!
            }, \qquad \forall u_{(j)} \in [0, 1], 
    \end{align*}
    which proves the first claim.
    
    Claim Two: let~$\cJ = \set{\zeta_1, \ldots, \zeta_c}$, s.t., $\zeta_1 < \cdots < \zeta_c < j$. 
    Following the notation of Theorem~\ref{theorem: conditional distribution of the order statistics}, we have $\ell = \zeta_c$.
    By Lemma~\ref{lemma: joint density of order statistics}, 
    \begin{align*}
        p \PAREN{ u_{(\cJ)} }  
            &= 
            m! \cdot \frac{
                \PAREN{u_{(\zeta_c)}}^{m - \zeta_c}
            }{
                \PAREN{m - \zeta_c}!
            } \cdot \frac{
                \PAREN{1 - u_{(\zeta_1)}}^{\zeta_1 - 1}
            }{
                \paren{\zeta_1 - 1}!
            } \cdot 
                \prod_{s = 2}^c \frac{
                    \PAREN{u_{(\zeta_{s - 1})} - u_{(\zeta_j)}}^{\zeta_{s} - \zeta_{s - 1} - 1}
                }{
                    \paren{\zeta_{s} - \zeta_{s - 1} - 1}!
                }
            ,\\
        p \PAREN{ u_{ ( \cJ \cup \set{j} ) } } 
            &= 
            m! \cdot \frac{
                \PAREN{u_{(j)}}^{m - j}
            }{
                \PAREN{m - j}!
            } \cdot \frac{
                \PAREN{u_{(\zeta_c)} - u_{(j)}}^{j - \zeta_c - 1}
            }{
                \PAREN{j - \zeta_c - 1}!
            } \cdot \frac{
                \PAREN{1 - u_{(\zeta_1)}}^{\zeta_1 - 1}
            }{
                \paren{\zeta_1 - 1}!
            } 
            \cdot 
                \prod_{s = 2}^c \frac{
                    \PAREN{u_{(\zeta_{s - 1})} - u_{(\zeta_{s})}}^{\zeta_{s} - \zeta_{s - 1} - 1}
                }{
                    \paren{\zeta_s - \zeta_{s - 1} - 1}!
                }.
    \end{align*}
    Hence, 
    \begin{align*}
        p \PAREN{ u_{(j)} \mid u_{(\cJ)} } 
            = \frac{p \PAREN{ u_{ ( \cJ \cup \set{j} ) } }}{p \PAREN{ u_{(\cJ)} }}
            = \frac{(m - \zeta_c)!}{ \paren{ j - \zeta_c  - 1}! \paren{m - j}!} \PAREN{
                    \frac{
                        u_{(\zeta_c)} - u_{(j)}
                    }{
                        u_{(\zeta_c)} 
                    }
                }^{j - \zeta_c  - 1} 
                \PAREN{
                    \frac{
                        u_{(j)} 
                    }{
                        u_{(\zeta_c)} 
                    }
                }^{m - j} 
                \frac{1}{u_{(\zeta_c)}}.
    \end{align*}
    
    Similarly, the densities of $p \PAREN{ u_{(\zeta_c)} }, p \PAREN{ u_{(\zeta_c)},  u_{(j)} }$ are given by 
    \begin{align*}
        p \PAREN{ u_{(\zeta_c)} }
            &= 
            m! \cdot \frac{
                \PAREN{u_{(\zeta_c)}}^{m - j}
            }{
                \PAREN{m - \zeta_c}!
            } \cdot \frac{
                \PAREN{1 - u_{(\zeta_c)}}^{\zeta_c - 1}
            }{
                \paren{\zeta_c - 1}!
            }\,, \\
        p \PAREN{ u_{(\zeta_c)},  u_{(j)} }
            &= 
            m! \cdot \frac{
                \PAREN{u_{(j)}}^{m - j}
            }{
                \PAREN{m - j}!
            } \cdot \frac{
                \PAREN{u_{(\zeta_c)} - u_{(j)}}^{j - \zeta_c - 1}
            }{
                \PAREN{j - \zeta_c - 1}!
            } \cdot \frac{
                \PAREN{1 - u_{(\zeta_c)}}^{\zeta_c - 1}
            }{
                \paren{\zeta_c - 1}!
            }\,.
    \end{align*}
    
    It is easy to see that 
    $$
        p \PAREN{ u_{(j)} \mid u_{(\zeta_c)} }
        =
        \frac{
            p \PAREN{ u_{(\zeta_c)},  u_{(j)} }  
        }{
            p \PAREN{ u_{(\zeta_c)} }
        } 
        = p \PAREN{ u_{(j)} \mid u_{(\cJ)} }.
    $$

    Claim Three: let~$\cJ = \set{\zeta_1, \ldots, \zeta_c}$, s.t., $\zeta_1 < \ldots < \zeta_c$, and there exists $c' \in [c - 1]$ for which $\zeta_{c'} < j < \zeta_{c' + 1}$.
    Following the notation of Theorem~\ref{theorem: conditional distribution of the order statistics}, we have $\ell = \zeta_{c'}, r = \zeta_{c' + 1}$.
    By Lemma~\ref{lemma: joint density of order statistics}, 
    \begin{align*}
        p \PAREN{ u_{(\cJ)} }  
            &= 
            m! \cdot \frac{
                \PAREN{u_{(\zeta_c)}}^{m - \zeta_c}
            }{
                \PAREN{m - \zeta_c}!
            } \cdot \frac{
                \PAREN{1 - u_{(\zeta_1)}}^{\zeta_1 - 1}
            }{
                \paren{\zeta_1 - 1}!
            } \cdot 
                \prod_{s = 2}^c \frac{
                    \PAREN{u_{(\zeta_{s - 1})} - u_{(\zeta_s)}}^{\zeta_s - \zeta_{s - 1} - 1}
                }{
                    \paren{\zeta_s - \zeta_{s - 1} - 1}!
                }
            \,,\\
        p \PAREN{ u_{ ( \cJ \cup \set{j} ) } } 
            &= 
            m! \cdot \frac{
                \PAREN{u_{(\zeta_c)}}^{m - \zeta_c}
            }{
                \PAREN{m - \zeta_c}!
            } \cdot \frac{
                \PAREN{1 - u_{(\zeta_1)}}^{\zeta_1 - 1}
            }{
                \paren{\zeta_1 - 1}!
            } \cdot \frac{
                    \PAREN{u_{(\zeta_{c'})} - u_{(j)}}^{j - \zeta_{c'} - 1}
                }{
                    \paren{j - \zeta_{c'} - 1}!
                }
            \cdot \frac{
                    \PAREN{u_{(j)} - u_{(\zeta_{c' + 1})}}^{\zeta_{c' + 1} - j - 1}
                }{
                    \paren{\zeta_{c' + 1} - j - 1}!
                }
            \\
            &\quad{} \cdot 
                \prod_{j = 2}^{c'} \frac{
                    \PAREN{u_{(\zeta_{s - 1})} - u_{(\zeta_s)}}^{\zeta_s - \zeta_{s - 1} - 1}
                }{
                    \paren{\zeta_s - \zeta_{s - 1} - 1}!
                }
            \cdot 
                \prod_{j = c' + 2}^{c} \frac{
                    \PAREN{u_{(\zeta_{s - 1})} - u_{(\zeta_s)}}^{\zeta_s - \zeta_{s - 1} - 1}
                }{
                    \paren{\zeta_s - \zeta_{s - 1} - 1}!
                }\,.
    \end{align*}
    Hence, 
    \begin{align*}
        p \PAREN{ u_{(j)} \mid u_{(\cJ)} } 
            &= \frac{p \PAREN{ u_{ ( \cJ \cup \set{j} ) } }}{p \PAREN{ u_{(\cJ)} }} 
            = \frac{
                    \paren{\zeta_{c' + 1} - \zeta_{c'} - 1}!
                }{
                    \paren{j - \zeta_{c'} - 1}! \paren{\zeta_{c' + 1} - j - 1}!
                } \\
            &\quad{} \cdot 
                \PAREN{\frac{u_{(\zeta_{c'})} - u_{(j)}}{u_{(\zeta_{c'})} - u_{(\zeta_{c' + 1})}}}^{j - \zeta_{c'} - 1} 
                \PAREN{\frac{u_{(j)} - u_{(\zeta_{c' + 1})}}{u_{(\zeta_{c'})} - u_{(\zeta_{c' + 1})}}}^{\zeta_{c' + 1} - j - 1}
                \frac{1}{u_{(\zeta_{c'})} - u_{(\zeta_{c' + 1})}}\,.
    \end{align*}
    
    Similarly, the densities of $p \PAREN{ u_{(\zeta_{c'})}, u_{(\zeta_{c' + 1})} }, p \PAREN{ u_{(\zeta_{c'})}, u_{(j)}, u_{(\zeta_{c' + 1})} }$ is given by 
    \begin{align*}
        p \PAREN{ u_{(\zeta_{c'})}, u_{(\zeta_{c' + 1})} }
            &= 
            m! \cdot \frac{
                \PAREN{u_{(\zeta_{c' + 1})}}^{m - \zeta_{c' + 1}}
            }{
                \PAREN{m - \zeta_{c' + 1}}!
            } \cdot \frac{
                \PAREN{u_{(\zeta_{c'})} - u_{(\zeta_{c' + 1})}}^{\zeta_{c' + 1} - \zeta_{c'} - 1}
            }{
                \PAREN{\zeta_{c' + 1} - \zeta_{c'} - 1}!
            } \cdot \frac{
                \PAREN{1 - u_{(\zeta_{c'})}}^{\zeta_{c'} - 1}
            }{
                \paren{\zeta_{c'} - 1}!
            }, \\
        p \PAREN{ u_{(\zeta_{c'})}, u_{(j)}, u_{(\zeta_{c' + 1})} }
            &= 
            m! \cdot \frac{
                \PAREN{u_{(\zeta_{c' + 1})}}^{m - \zeta_{c' + 1}}
            }{
                \PAREN{m - \zeta_{c' + 1}}!
            } \cdot \frac{
                \PAREN{u_{(\zeta_{c'})} - u_{(j)}}^{j - \zeta_{c'} - 1}
            }{
                \PAREN{j - \zeta_{c'} - 1}!
            } \\
            &\quad{} \cdot \frac{
                \PAREN{u_{(j)} - u_{(\zeta_{c' + 1})}}^{\zeta_{c' + 1} - j - 1}
            }{
                \PAREN{\zeta_{c' + 1} - j - 1}!
            } \cdot \frac{
                \PAREN{1 - u_{(\zeta_{c'})}}^{\zeta_{c'} - 1}
            }{
                \paren{\zeta_{c'} - 1}!
            }.
    \end{align*}
    
    It is easy to see that 
    $$
        p \PAREN{ u_{(j)} \mid u_{(\zeta_{c'})}, u_{(\zeta_{c' + 1})} } 
        = \frac{
            p \PAREN{ u_{(\zeta_{c'})}, u_{(j)}, u_{(\zeta_{c' + 1})} }
        }{
            p \PAREN{ u_{(\zeta_{c'})}, u_{(\zeta_{c' + 1})} } 
        } 
        =  p \PAREN{ u_{(j)} \mid u_{(\cJ)} }. 
    $$
\end{proof}

\section{Proofs for Section~\ref{sec: lower bounds}}

\subsection{Proofs for Section~\ref{subsec: random access}}
\label{appendix: proof for subsec: random access}

\begin{proof}[Proof of Theorem~\ref{theorem: lower bound for random access}]
    We do not specify the family~$\cH$ and the distribution~$\mu$ on~$\cH$ directly. 
    Instead, we show how we can sample a~$\hist$ from~$\cH$ according to~$\mu$: for each~$i \in [m]$, independently set 
    \begin{equation} \label{equa: hist dist for rnd access lower bound}
        \hist[i] \doteq 
        \begin{cases}
            n,  & \text{ w.p. } \frac{2k}{m}\,, \\
            0,  & \text{ w.p. } 1 - \frac{2k}{m}\,. 
        \end{cases}
    \end{equation}
    This can also be understood as, for each item~$i \in [m]$, with probability~$2 k / m$ all of the~$n$ clients votes for~$i$; and with probability~$1 - 2 k / m$, none of the clients votes for~$i$.

    Assume that: 1)~$\cA$ retrieves entries from~$\hist$ (via random access) without repetition.
    This only decreases its access cost, since accessing a previously encountered entry does not provide additional information.
    2) If~$\cA$ terminates before retrieving all entries in~$\hist$, it continues to read the remaining entries without being charged for the additional accesses.
    This enables~$\cA$ to obtain more entries for free.
    Now, let~$J_1, \ldots, J_m \in [m]$ be the order in which~$\cA$ accesses the entries.
    The sequence constitutes a permutation of~$[m]$, and for each~$t \in [m]$, the choice of~$J_t$ can depend on previous choices~$J_1, \ldots, J_{t - 1}$ and outcomes~$\hist[J_1], \ldots, \hist[J_{t - 1}]$.
    However, whatever the choice of~$J_t$ is, the distribution of~$\hist[J_t]$ (conditioned on previous choices and outcomes) is till given by Equation~(\ref{equa: hist dist for rnd access lower bound}). 
    Hence,~$\hist[J_1], \ldots, \hist[J_{m}]$ can be viewed as independent random variables. 
    
    Consider the following events. 
    Event~$E_1:$~$\hist$ has at least~$k$ non-zero entries.
    Since each entry of~$\hist$ is generated independently, 
    via Chernoff bound (Fact~\ref{fact: chernoff bound sampling with replacement}),
    $$
        \P{ \bar{E_1} } 
            \le \exp \PAREN{ - \frac{ \paren{0.5}^2 \cdot 2 k}{2} }
            \le \exp \paren{-\paren{0.5}^2} \le 0.78\,. 
    $$
    Event~$E_2:$ the number of non-zero entries among~$\hist[J_1]$ $,\ldots, \hist[J_{m / 50}]$ is less than~$k$.
    Since~$\hist[J_1], \ldots, \hist[J_{m / 50}]$ are  independent, 
    via Chernoff bound (Fact~\ref{fact: chernoff bound sampling with replacement}), and noting that~$k \ge 1$, 
    $$
        \P{ \bar{E_2} } \le \PAREN{ \frac{ e^{24} }{ 25^{25} } }^{k / 25} 
        \le \PAREN{ \frac{ e^{24} }{ 25^{25} } }^{1 / 25} \le 0.11\,.
    $$
    Event~$E_3:$~$\cA$ returns an~$(n - 1, k)$-accurate solution.
    By assumption, we have~$\P{\bar{E_3}} \le \beta < 0.1$.
    
    Hence,~$\P{ E_1 \cap E_2 \cap E_3} \ge 1 -  \P{ \bar{E_1} } -  \P{ \bar{E_2} } -  \P{ \bar{E_3} } \in \Omega(1)$.
    Observe that, when~$E_1$ and~$E_3$ happens, each item~$i$ returned by~$\cA$ must have frequency~$\hist[i] = n$. 
    However, when~$E_2$ happens,~$\cA$ cannot see~$k$ items with non-zero frequency, from its first~$m / 50$ retrievals. 
    
    It follows that , with probability at least~$\P{ E_1 \cap E_2 \cap E_3}$,~$\cA$ has access cost at least~$m / 50$, which proves the theorem. 

\end{proof}

\subsection{Proofs for Section~\ref{subsec: sorted access}}
\label{appendix: proof for subsec: sorted access}

\begin{proof}[Proof of Lemma~\ref{lemma: probability of reporting back candiates}]

    Let~$\cO \doteq \binom{[m]}{k}$ be a shorthand for the collection of all possible outputs of~$\cA$. 
    Observe that, given an input histogram~$\hist_S$ or~$\hist_{{\HighFreqSet, \LowFreqSet}}$, if~$\cA$'s output is~$(n - 2, k)$-accurate, then it must be a subset of~$S$ of size~$k$.
    Therefore, define~$\cG \doteq \binom{S}{k}$ be a shorthand for the collection of all~$(n - 2, k)$-accurate outputs.
    Via the assumption that~$\cA$ outputs an~$(n - 2, k)$-accurate solution with probability at least~$1 - \beta$, it holds that
    $$
        \P{ \cA(\hist_S) \in \cG } \ge 1 - \beta\,. 
    $$
    
    Let~$\cF \doteq \binom{S}{|S| / k }$ be a shorthand for the collection all possible outcomes of~$\LowFreqSet$.
    For each~$s_\ell \, \in \cF$, define
    ~$\cO_{s_\ell} \doteq \set{ o \in \cO : o \cap s_\ell \neq \emptyset }$, 
    the collection of sets in~$\cO$ that has nonempty intersection with~$s_\ell$, 
    and~$\cG_{s_\ell} \doteq \set{ o \in \cG : o \cap s_\ell \neq \emptyset }$, 
    the collection of sets in~$\cG$ that has nonempty intersection with~$s_\ell$.
    Conditioned on~$\LowFreqSet = s_\ell$, 
    \begin{align*}
        &\P{ 
            \cA( \hist_{{\HighFreqSet, \LowFreqSet}} ) 
            \cap \LowFreqSet 
            \neq \emptyset 
            \mid \LowFreqSet = s_\ell
        } 
        = 
        \P{ 
            \cA( \hist_{{\HighFreqSet, \LowFreqSet}} ) 
            \in 
            \cO_{s_\ell}
            \mid \LowFreqSet = s_\ell
        } 
        \ge 
        \P{ 
            \cA( \hist_{{\HighFreqSet, \LowFreqSet}} ) 
            \in 
            \cG_{s_\ell}
            \mid \LowFreqSet = s_\ell
        },
    \end{align*}
    where the inequality holds as~$\cO_{s_\ell} \supseteq \cG_{s_\ell}$. 
    Moreover, since~$\cA$ is~$\paren{\eps, \delta}$-DP, it holds that
    \begin{align}
        \label{ineq: dp inequality}
        \P{ 
            \cA( \hist_{{\HighFreqSet, \LowFreqSet}} ) 
            \in 
            \cG_{s_\ell}
            \mid
            \LowFreqSet = s_\ell
        }
        \ge
        \frac{1}{e^{\eps}} \paren{  \P{ \cA(\hist_S) \in \cG_{s_\ell} } - \delta }\,.
    \end{align}
        Further, since the events that~$\cA(\hist_S) = o$ are mutually exclusive for different values of~$o \in \cG$, we have
    \begin{align}
        \P{ \cA(\hist_S) \in \cG }
            &=  \sum_{o \in \cG} 
             \P{ \cA(\hist_S) = o } \ge 1 - \beta\,,
        \\
        \P{ \cA(\hist_S) \in \cG_{s_\ell} }
            &=  \sum_{o \in \cG, o \cap s_\ell \neq \emptyset} 
             \P{ \cA(\hist_S) = o } 
            = \sum_{o \in \cG} 
            \indicator{o \cap s_\ell \neq \emptyset} \cdot \P{ \cA(\hist_S) = o }\,.
    \end{align}
    Finally, 
    \begin{align*}
            \P{\LowFreqSet, \cA}{ 
                \cA( \hist_{{\HighFreqSet, \LowFreqSet}} ) 
                \cap \LowFreqSet 
                \neq \emptyset 
            }
        &=
        \sum_{s_\ell \, \in \cF} \P{\LowFreqSet = s_\ell} \cdot \P{ 
            \cA( \hist_{{\HighFreqSet, \LowFreqSet}} ) 
            \in 
            \cO_{s_\ell}
            \mid
            \LowFreqSet = s_\ell
        } \\
        &\ge
        \sum_{s_\ell \, \in \cF}  \P{\LowFreqSet = s_\ell} \cdot \P{ 
            \cA( \hist_{{\HighFreqSet, \LowFreqSet}} ) 
            \in 
            \cG_{s_\ell}
            \mid
            \LowFreqSet = s_\ell
        } \\
        &\ge \sum_{s_\ell \, \in \cF} 
            \P{\LowFreqSet = s_\ell} \cdot e^{-\eps} \left( 
                \sum_{o \in \cG} \indicator{ o \cap s_\ell \neq \emptyset} \cdot \P{ \cA(\hist_S) = o }
                - \delta
            \right) \\
        &= - e^{-\eps} \delta + e^{-\eps} \sum_{o \in \cG} 
            \P{ \cA(\hist_S) = o } \left(
                \sum_{s_\ell \, \in \cF} \indicator{ o \cap s_\ell \neq \emptyset} \cdot 
                \P{\LowFreqSet = s_\ell}
            \right) \\
        &\stackrel{(a)}{=} - e^{-\eps} \cdot \delta + e^{-\eps} \sum_{o \in \cG} 
            \P{ \cA(\hist_S) = o } \PAREN{
                1 - \frac{
                    \binom{|S| - k}{|S| / k} 
                }{
                    \binom{|S|}{|S| / k}
                }
            } \\
        &\stackrel{(b)}{\ge} - e^{-\eps} \cdot \delta + e^{-\eps} \sum_{o \in \cG} 
            \P{ \cA(\hist_S) = o } \PAREN{
                1 - \exp \paren{ - 1 }
            } \\
        &\ge - e^{-\eps} \cdot \delta + e^{-\eps} \PAREN{1 - \beta} \PAREN{
                1 - e^{-1}
            } \\
        &\ge e^{-\eps} \PAREN{1 - \beta - \delta - e^{-1}}\,,
    \end{align*}
    where equation~$(a)$ follows, since~$\sum_{s_\ell \, \in \cF} \indicator{ o \cap s_\ell \neq \emptyset} \cdot \P{\LowFreqSet = s_\ell}$ can interpreted as the probability that given a subset~$o$ of~$S$ of size~$k$, the sampled subset~$\LowFreqSet$ has nonempty intersection with~$o$;
    and inequality~$(b)$ follows, since 
    \begin{align*}
        \frac{
            \binom{|S| - k}{|S| / k} 
        }{
            \binom{|S|}{|S| / k}
        }
        &= \frac{|S| - k}{|S|} \cdots \frac{|S| - k - |S| / k + 1}{|S| - |S| / k + 1} \\
        &\le \PAREN{
            \frac{|S| - k}{|S|}
        }^{|S| / k} \\
        &\le \exp \PAREN{
            - \frac{k}{|S|} \cdot \frac{|S|}{k}
        } \\
        &= \exp \paren{ - 1 }\,.
    \end{align*}
\end{proof}

\subsection{Proofs for Section~\ref{subsec: random sorted and sorted access}}
\label{appendix: proof for subsec: random sorted and sorted access}

\begin{proof}[Proof of Lemma~\ref{lemma: lower bound of E1}]
        Assume that: 1) if~$\cA$ terminates before performing~$\eta$ accesses operations, it continues to perform more until it reaches~$\eta$, and will not be charged for any additional access.
    Now, let~$J_1, \ldots, J_{\eta}$ represents the first~$\eta$ operations of~$\cA$: each~$J_t$ is either a character `s', implying that the~$t^{(th)}$ operation is a sorted access, or an integer in~$[m]$, implying that the~$t^{(th)}$ operation is a random access to the entry~$\hist_{\HighFreqSet, \LowFreqSet}[J_t]$.
    Further, let~$J_1', \ldots, J_t'$ be the subsequence of all random access in~$J_1, \ldots, J_{\eta}$.
    
    Consider an alternative algorithm~$\cA'$ that operates as follows: it first performs~$9 \eta$ sorted accesses to obtain all frequencies of all items from~$\HighFreqSet$, followed by random accesses~$J_1', \ldots, J_t'$. 
    It is clear that,~$\cA'$ retrieves a greater number of entries than the first~$\eta$ operations of~$\cA$. 
    
    We bound the probability that~$\cA'$ does not retrieve an entry from~$\LowFreqSet$.
    First consider random access~$J_1'$.
    If~$J_1'$ had already been retrieved by sorted access, then clearly~$J_1' \notin \LowFreqSet$. 
    Otherwise, conditioned on the fact that the entries~$\HighFreqSet$ have been determined, according to the manner~$S$ and~$\LowFreqSet$ are generated,
    each item in~$[m] \setminus \PAREN{\HighFreqSet}$ belongs to~$\LowFreqSet$ with equal probability.
    Hence,
    $$
        \P{J_1' \notin \LowFreqSet} 
            \ge 1 - \frac{\tau / k}{m - (k - 1)\tau / k}\,.
    $$
    
    In general, for each~$1 < \ell \le t$, suppose that~$J_1', \ldots, J_{\ell - 1}'$ does not belong to~$\LowFreqSet$, and has revealed~$m_{\ell} \le \ell - 1$ distinct items from~$[m] \setminus \PAREN{\HighFreqSet}$.
    Whatever the choice of~$J_\ell'$ is, and the items~$J_1', \ldots, J_{\ell - 1}'$ are, each of the remaining items that have not been queried belong to~$\LowFreqSet$ with equal probability.
    Hence
    $$
        \P{J_\ell' \notin \LowFreqSet} 
            \ge 1 - \frac{\tau / k}{m - (k - 1)\tau / k - m_\ell}.
    $$
    
    Noting that~$m_\ell < t \le \eta$, we have
    \begin{align*}
        \P{J_1', \ldots, J_t' \notin \LowFreqSet}
            &\ge \prod_{\ell = 1}^t \PAREN{
                1 - \frac{\tau / k}{m - (k - 1)\tau / k - m_\ell}
            } \\
            &\ge \prod_{\ell = 1}^t \PAREN{
                1 - \frac{ \tau / k }{m - (k - 1)\tau / k - \eta} 
            } \\
            &\stackrel{(a)}{\ge} \prod_{\ell = 1}^t \exp \PAREN{
                - \frac{2 \tau / k}{m - (k - 1)\tau / k - \eta}
            } \\
            &\ge \exp \PAREN{
                - \frac{ \tau^2 / (10 k) }{m - (k - 1)\tau / k - \eta} 
            } \\
            &\ge \exp \PAREN{
                - \frac{ m / 10 }{m - 21 \sqrt{mk} / 20} 
            } \\
            &\ge \exp \PAREN{
                - \frac{ m / 10 }{m - 21 m / 40} 
            } \\
            &\ge 0.81.
    \end{align*}
    where inequality~$(a)$ holds, since~$1 - x \ge e^{-2x}, \forall x \in [0, 3 / 4]$, and
    \begin{align}
        \tau / k \le \frac{3}{4} \PAREN{m - (k - 1)\tau / k - \eta} \\
        \Longleftrightarrow 
        \tau \PAREN{ \frac{3}{4} + \frac{1}{4k} + \frac{3}{40} } \le \frac{3}{4} m 
    \end{align}
    The last inequality holds, since~$\frac{3}{4} + \frac{1}{4k} + \frac{3}{40} \le 3 / 2$ for~$k \in \N^+$, and 
    $2 \tau = 2 \sqrt{m k} \le m$, i.e.,~$k \le m / 4$. 
    \end{proof}